\documentclass[apj]{emulateapj}
\usepackage{amsmath}
\usepackage{graphicx}
\usepackage{epsfig}
\usepackage{float}
\usepackage{amssymb}
\usepackage{subfigure}
\usepackage{natbib}
\newcommand{\az}{_{A,Z}}
\newcommand{\midtilde}{\raisebox{-0.25\baselineskip}{\textasciitilde}}

\shorttitle{New equations of state in core-collapse supernova simulations}
\shortauthors{Hempel et al.}

\begin{document}

\title{New equations of state in simulations of core-collapse supernovae}

\author{M.~Hempel$^{1}$,
T.~Fischer$^{2,3}$,
J.~Schaffner-Bielich$^{4}$
and 
M.~Liebend{\"o}rfer$^{1}$}

\affil{
  $^1$Departement Physik, Universit\"at Basel, Klingelbergstr.~82, 4056 Basel,
Switzerland  \\
  $^2$GSI, Helmholtzzentrum f\"ur Schwerionenforschung GmbH, Planckstr.~1, 64291
Darmstadt, Germany\\
  $^3$Institut f\"ur Kernphysik, Technische Universit{\"a}t Darmstadt, Schlossgartenstr.~9, 64289 Darmstadt,
Germany \\
  $^4$Institut f\"ur Theoretische Physik, Ruprecht-Karls-Universit\"at,
Philosophenweg 16, 69120 Heidelberg, Germany
}

\begin{abstract}
We discuss three new equations of state (EOS) in core-collapse supernova simulations. The new EOS are based on the nuclear statistical equilibrium model of Hempel and Schaffner-Bielich (HS), which includes excluded volume effects and relativistic mean-field (RMF) interactions. We consider the RMF parameterizations TM1, TMA, and FSUgold. These EOS are implemented into our spherically symmetric core-collapse supernova model, which is based on general relativistic radiation hydrodynamics and three-flavor Boltzmann neutrino transport. The results obtained for the new EOS are compared with the widely used EOS of H.~Shen et al.\ and Lattimer \& Swesty. The systematic comparison shows that the model description of inhomogeneous nuclear matter is as important as the parameterization of the nuclear interactions for the supernova dynamics and the neutrino signal. Furthermore, several new aspects of nuclear physics are investigated: the HS EOS contains distributions of nuclei, including nuclear shell effects. The appearance of light nuclei, e.g., deuterium and tritium is also explored, which can become as abundant as alphas and free protons. In addition, we investigate the black hole formation in failed core-collapse supernovae, which is mainly determined by the high-density EOS. We find that temperature effects lead to a systematically faster collapse for the non-relativistic LS EOS in comparison to the RMF EOS. We deduce a new correlation for the time until black hole formation, which allows to determine the maximum mass of proto-neutron stars, if the neutrino signal from such a failed supernova would be measured in the future. This would give a constraint for the nuclear EOS at finite entropy, complementary to observations of cold neutron stars.
\end{abstract}

\journalinfo{Published in ApJ, 748, 70 (2012)}

\maketitle

\section{Introduction}
Supernova explosions of stars more massive than 8~M$_\odot$ are an active
subject of research in astrophysics. The core-collapse supernova problem is
related to the revival of the stalled bounce shock. It forms when the collapsing
stellar core bounces back above normal nuclear matter density. Explosions in
spherical symmetry have been obtained only for the low mass 8.8~M$_\odot$
O--Ne--Mg core from \citet{Nomoto:1983,Nomoto:1984,Nomoto:1987} by
\citet{Kitaura:etal:2006} and \citet{Fischer:etal:2010a}. More massive
progenitor stars have extended high-density Si-layers surrounding the central iron
core. The post-bounce evolution leads to a mass accretion period that
lasts for several hundred milliseconds. For such iron-core progenitor stars
explosions do not occur in spherically symmetric simulations. The bounce shock
continuously looses energy by dissociation of heavy nuclei and neutrino heating
is not sufficient enough to revive the standing bounce shock.

In addition to the standard scenario of neutrino-driven explosions (see
\citet{BetheWilson:1985}), several alternative explosion mechanisms have been
proposed. These are the magneto-rotational mechanism by
\citet{LeBlankWilson:1970} and the acoustic mechanism by \citet{Burrows:2006a}.
All of which are working in multiple spatial dimensions. Multi-dimensional
core-collapse supernova models, based on sophisticated neutrino transport
approximations and general relativity, have become available only recently.
These models have shown to increase the neutrino heating efficiency (see, e.g.,
\citet{Miller:etal:1993}, \citet{Herant:etal:1994}, \citet{JankaMueller:1996},
\citet{Burrows:2006b}) and help to aid the understanding of aspherical
explosions (see, for example, \citet{Bruenn:etal:2006}; \citet{MarekJanka:2009};
and \citet{Muller:etal:2010}).

Apart from the dimensionality, the simulations are also affected by
uncertainties in the nuclear physics involved, which can be divided in weak
processes and the equation of state (EOS). Changes in the nuclear physics input
can have dramatic consequences and can even generate explosions in spherically
symmetric models as was illustrated recently by \citet{Sagert:etal:2009} and
\citet{Fischer:2011}. The authors explored the state of matter around and above
nuclear matter density and investigated the possibility of the quark-hadron
phase transition during the early post-bounce evolution of low and intermediate
mass iron-core progenitor stars. It was found that the phase transition leads to
the formation of a strong hydrodynamic shock wave which triggers the explosion.
In the present article we study new aspects of the hadronic EOS and their
consequences for the supernova dynamics.

Until today, there are only a few hadronic EOS available which cover a
sufficiently large domain in density, temperature, and electron fraction so that
they can be used in simulations of core-collapse supernovae.  
Usually, in such simulations the EOS is implemented in form of a table. The most prominent among
them are the two ``classic'' EOS from \citet{LattimerSwesty:1991} (LS) and
H.~\citet{Shen:etal:1998,shen98} (STOS) which are commonly used in computational
astrophysics.
The LS EOS is provided in form of programming routines 
for three different values of the nuclear incompressibility $K$ of 180, 220, and
375~MeV. STOS is an EOS in tabular form which was calculated for the relativistic mean-field (RMF) model
TM1 \citep[][]{SugaharaToki:1994}. There has been a lot of progress for the
supernova EOS in the last years. Recently, the new hadronic EOS tables of
G.~\citet{shen2011a,shen2011b} became available, which are based on the virial
expansion and two different RMF interactions in the Hartree approximation. In their first article they
applied NL3 (G.~\citet{shen2011a}), followed by a table for FSUgold
(G.~\citet{shen2011b}). Furthermore also an update of the original STOS EOS
table appeared which has improved resolution, accuracy and grid-spacing
(H.~\citet{hshen2011}). In this article also another table was presented in which
lambda hyperons are taken into account. Similarly, several other extensions of the 
STOS table are available, either with the formation of hyperons \citep{ishizuka08} or
the inclusion of the QCD phase transition
\citep{Nakazato:etal:2008a,Sagert:etal:2009,Fischer:2011}. In the present
investigation we apply the new EOS tables of
\citet{HempelSchaffnerBielich:2010}.

An important aspect of the supernova EOS is the formation of light nuclei and
their properties in the hot and dense medium. Note that the two classic
supernova EOS include only alpha particles of all possible light nuclei, which
are implemented with excluded volume effects. \citet{Typel:etal:2010} use a
generalized RMF and a quantum statistical model \citep{roepke11} to study supernova matter taking
into account the most important light nuclei with mass number $A\leq 4$.
Significant differences in the composition were found compared to STOS, similar
as in the recent studies of \citet{HorowitzSchwenk:2006},
\citet{Oconnnor:etal:2007}, \citet{SumiyoshiRoepke:2008},
\citet{Heckel:etal:2010}, and \citet{HempelSchaffnerBielich:2010}. Also
thermodynamic variables, like ,e.g., the symmetry energy, are modified due to the
appearance of light nuclei. A comparison between the excluded volume approach with all light nuclei, which is also used in the HS EOS, and the quantum many-body approaches of \citet{roepke11} and \citet{Typel:etal:2010} is given by \citet{hempel11b}. It is found that the dissolution of light nuclei happens at similar densities and that the
excluded volume approach gives an acceptable representation of the real
microscopic medium effects, like, e.g., Pauli-blocking. Recently, the importance of light nuclei in
supernova matter was also shown by a heavy-ion collision experiment. In the isoscaling
analysis of \citet{Natowitz:etal:2010} the measured symmetry energy can only be
explained, if the formation of light nuclei is taken properly into account.
Quite remarkably, the deduced temperatures and densities correspond to
conditions which are typical for matter in core-collapse supernovae. In the
supernova environment light nuclei can possibly influence the neutrino transport and
consequently the supernova neutrino signal and dynamics, see, e.g.,
\citet{Oconnnor:etal:2007}, \citet{Arcones:etal:2008}, \citet{SumiyoshiRoepke:2008},
and \citet{Nakamura:etal:2009}. However, so far there are no investigations of
core-collapse supernovae, which consistently take into account all light nuclei.
With this article we further advance in this direction, by applying the EOS
tables of the statistical model of \citet{HempelSchaffnerBielich:2010} (HS) to
simulations of massive stars.

Statistical models are characterized by having a distribution of nuclei, in
which light nuclei can also be included. On the other hand, the complexity and
variety of the involved aspects of nuclear physics require some modeling.
Important aspects in the supernova context are for example the interactions of
unbound nucleons, excluded volume effects, excited states, Coulomb interactions,
nuclear binding energies, shell effects or surface modifications of nuclei, see,
e.g.,
\citet{Ishizuka:etal:2003}, \citet{BovinaMishustin:2004}, \citet{NadyozhinYudin:2004}, \citet{NadyozhinYudin:2005}, \citet{Souza:etal:2009}, \citet{Blinnikov:etal:2009},
\citet{HempelSchaffnerBielich:2010}, \citet{BotvinaMishustin:2010}, \citet{Arcones:etal:2010}, \citet{RadutaGulminelli:2010}, \citet{yudin11}, and \citet{furusawa11}. Interestingly, similar statistical
models are also used for the analysis of multifragmentation experiments
\citep{koonin87,gross90,bondorf95}. In this article we apply the new EOS from
\citet{HempelSchaffnerBielich:2010}. The HS EOS uses nuclear statistical
equilibrium (NSE) with excluded volume effects for distributions of nuclei and RMF interactions of unbound
nucleons. In this model all possible light nuclei (e.g., deuterium and tritium)
and heavy nuclei up to mass number $A\sim 330$ from the neutron to the
proton dripline are included. The knowledge of the detailed composition makes it
possible to analyze the impact of different baryon contributions during the
different phases of core-collapse supernovae. 

The nuclear distributions are especially relevant for weak reactions. For
example the new treatments of electron captures on heavy nuclei by
\citet{Langanke:etal:2003} and \citet{Hix:etal:2003} and inelastic
neutrino--nucleon (nuclei) scattering by \citet{Langanke:etal:2008} are all based
on NSE distributions of nuclei. Idealistically, the nuclear physics input in the
weak reactions should be as consistent as possible with the EOS and therefore
the two parts should be built from the same degrees of freedom. However, also in
the present study we use additional simplifications, which will be specified
below.

To classify the characteristic features of HS, we take the two aforementioned
classic supernova EOS, i.e., LS and STOS, as standard references. These two EOS
both use the single nucleus approximation (SNA), i.e., the distribution of heavy
nuclei is replaced by a single representative nucleus. \citet{burrows84} showed that
the SNA has only a minor effect on thermodynamic variables. However, using the
SNA instead of nuclear distributions represents a significant difference for the
weak reactions, as mentioned already above. \citet{Souza:etal:2009} found
that the SNA gives also a systematically larger representative heavy nucleus compared to the average
of the distribution. There are further limitations of the description of nuclei
in the two EOS models: STOS is based on the Thomas--Fermi and local-density
approximation and a minimization of the free energy for parameterized
nucleon-density profiles within an RMF model. LS is based on a non-relativistic
liquid drop description including surface effects. Both models do not include
nuclear shell effects and only give an approximate description of the iron-group nuclei,
which appear at low temperatures. This leads to artificial shifts in the EOS at
the transition to non-NSE. Contrary, due to the use of experimental nuclear masses, the HS
model reproduces the thermodynamic state of the ideal gas of, e.g., $^{56}$Fe or
$^{56}$Ni (depending on the proton-to-baryon ratio) at low temperatures by
construction and naturally includes shell effects. Thus, it represents a major
improvement to other commonly used EOS as it gives a more detailed description
of the chemical composition and the nuclear effects involved at low densities
below $\sim 10^{12}$ g/cm$^3$.

However, with statistical models it is difficult to describe the transition to
uniform nuclear matter which occurs around half times saturation density, i.e., around $10^{14}$~g/cm$^3$.
Microscopic SNA models allow a more detailed description of the effects which
occur at such large densities, like the formation of (non-spherical) inhomogeneous
structures, the so-called nuclear pasta phases, see,
e.g., \citet{ravenhall83} and \citet{newton09}. In \citet{HempelSchaffnerBielich:2010} it was
shown that the thermodynamic variables in the transition region calculated with
the HS model are in satisfactory agreement with the results of STOS and LS. From
this point of view, the present investigation serves as a practical test whether
the HS EOS is suitable to be used for all possible conditions in core-collapse
supernova simulations. 

Recently, five new supernova EOS tables were calculated with the HS model for
different nuclear interactions and are available online.\footnote{See
\texttt{http://phys-merger.physik.unibas.ch/\midtilde
hempel/eos.html}\label{eospage}.} One of the new HS EOS tables is based on the
same parameterization TM1 \citep[][]{SugaharaToki:1994} of RMF interactions of
the nucleons as STOS. This makes it possible to investigate the impact of the model
assumptions for the description of inhomogeneous nuclear matter below saturation
density, which is one of the main topics of the present study. In addition to
TM1 and the general features of the HS model, we also investigate two other RMF
parameterizations of nuclear interactions, TMA (\citet{Toki:etal:1995}) and
FSUgold (\citet{ToddRutelPiekarewicz:2005, Piekarewicz:2007}). We note that
these two interactions have not been applied in supernova-simulations so far.
The different nuclear interactions become significant at high densities and we
analyze to which extent they can influence the supernova dynamics and the
observable neutrino signal. For the distinction of the different HS
EOS we append the used parameterization in brackets, e.g., ``HS~(TMA)''. To be
explicit, sometimes we also use ``STOS~(TM1)'' although the STOS table exists
only for TM1. For the LS EOS we append the used incompressibility in brackets,
e.g., ``LS (180).''

The manuscript is organized as follows. In \S \ref{sec_theory}, we present our
core-collapse model, including the new HS EOS. The results from simulations of
the collapse, bounce, and the early post-bounce phases of a 15~M$_{\odot}$
progenitor are presented in \S \ref{sec_15}. We compare HS (TM1)
with the LS (180) and the STOS (TM1) EOS and also analyze the neutrino signal.
In \S \ref{sec_40}, we discuss the neutrino signal of a 40~M$_{\odot}$
progenitor. For this massive progenitor we apply five different EOS: LS (180),
LS (220), STOS (TM1), HS (FSUgold), and HS (TMA). We focus on the time until
black hole formation, and investigate its relation to characteristic properties
of the different nuclear interactions, before we close with a summary in \S
\ref{sec_sum}.
\section{Theoretical and numerical setup}
\label{sec_theory}
\subsection{Core-collapse Supernova Model}

\begin{table}[tb]
\centering
\caption{Neutrino Reactions Considered, Including References}
\begin{tabular}{ c c }
\hline
\hline
Weak Process
\footnote{$\nu=\{\nu_e,\bar{\nu}_e,\nu_{\mu/\tau},\bar{\nu}_{\mu/\tau}\}$
and $N=\{n,p\}$.}
&
References
\\
\hline
$\nu_e + n \rightleftarrows p + e^-$
&
\citet{Bruenn:1985}\\
&\\
$\bar{\nu}_e + p \rightleftarrows n + e^+$
&
\citet{Bruenn:1985}\\
&\\
$\nu_e + \left<A\right> \rightleftarrows \left<A\right> + e^-$
&
\citet{Bruenn:1985}\\
&\\
$\nu + N/\langle A \rangle/\alpha \rightleftarrows \nu' + N/\langle A
\rangle/\alpha$
&
\citet{Bruenn:1985};\\
&\citet{MezzacappaBruenn:1993a}\\
&\\
$\nu + e^\pm \rightleftarrows \nu' + e^\pm$
&
\citet{Bruenn:1985};\\
&
\citet{MezzacappaBruenn:1993c}\\
&\\
$\nu + \bar{\nu} \rightleftarrows e^- + e^+$
&
\citet{Bruenn:1985};\\
&\citet{MezzacappaMesser:1999}\\
&\\
$\nu + \bar{\nu} + N + N \rightleftarrows N + N$
&
O.~E.~B.~Messer\footnote{Private communications,
based on \citet{HannestadRaffelt:1998}.} \\
& \\
$\nu_e + \bar\nu_e \rightleftarrows \nu_{\mu/\tau} + \bar\nu_{\mu/\tau}$
&
\citet{Buras:etal:2003}; \\
&
\citet{Fischer:etal:2009} \\
\hline
\hline
\end{tabular}
\flushleft
{\bf Notes:}
\label{tab:nu-reactions}
\end{table}
Our core-collapse model was originally constructed based on Newtonian radiation
hydrodynamics and three flavor Boltzmann neutrino transport \citep[for details
see][]{MezzacappaBruenn:1993a,MezzacappaBruenn:1993b,MezzacappaBruenn:1993c}.
It was extended to solve the general relativistic equations for both
hydrodynamics and neutrino transport in \citet{Liebendoerfer:etal:2001a}, based
on the following line-element
\begin{equation*}
ds^2 = \alpha^2dt^2 + \left(\frac{r'}{\Gamma}\right)^2da^2
+ r^2d\Omega
\end{equation*}
in co-moving coordinates system time $t$ and enclosed baryon mass $a$, with
metric coefficients lapse function $\alpha(t,a)$ and $\Gamma(t,a)$, where
$r'=\partial r/\partial a$. $r^2 d\Omega$ describes a 2-sphere, in spherical
coordinates $(\theta,\phi)$, of radius $r(t,a)$. Equations for momentum and
energy conservation are obtained via the co-variant derivative of the
stress-energy tensor, $\nabla_i\,T^{\rm ik}=0$, with
\begin{eqnarray*}
T^{\rm tt} &=& \rho(1+e+J), \,\,\,\,
T^{\rm aa} = p + \rho K, \,\,\,\, 
T^{\rm at} = T^{\rm ta} = \rho H, \\
T^{\theta\theta} &=& T^{\phi\phi} = p + \frac{1}{2}\rho(J-K),
\end{eqnarray*}
where $\rho$ and $p$ are rest-mass density (further discussed below) and matter
pressure. $e$ is the specific internal matter energy with respect to the
rest-mass density. $J$ and $K$ are the neutrino contributions to the specific
internal energy and specific pressure. $H$ is the specific energy exchange
between matter and the neutrino radiation field. $H, J$, and $K$ are the neutrino
moments and are given by momentum integrals of the neutrino distribution
functions \citep[for details see][]{Liebendoerfer:etal:2004}. Furthermore,
special attention has been devoted to accurately conserve energy, momentum, and
lepton number in \citet{Liebendoerfer:etal:2004}.

Neutrino contributions, more precisely the evolution of the neutrino
distribution functions $f_\nu$, are computed solving the Boltzmann transport
equation for ultrarelativistic and mass-less fermions, for each flavor $\nu=\nu_e, \bar \nu_e, \nu_{\mu/\tau}, \bar \nu_{\mu/\tau}$. In
addition to the space-time coordinates, it depends on the momentum angle
$\theta$ and energy $E$. The left-hand side of the Boltzmann equation determines
the phase-space derivative due to transport, and the right-hand side takes
collisions into account. The latter relates to the weak processes considered,
which are listed in Table~\ref{tab:nu-reactions}. Note that for the
alpha particles only neutral--current elastic scattering reactions are included. The emission of
$(\mu/\tau)$-neutrino pairs via the annihilation of trapped
$(\nu_e,\bar{\nu}_e)$ has been implemented into our model in
\citet{Fischer:etal:2009}. In the Boltzmann transport representation, for
scattering processes we employ reaction kernels, e.g.,  they take into account
final state blocking and depend on incoming and outgoing neutrino angle and
energy. For charged-current processes, we employ reaction rates which depend
only on the incoming neutrino energy.

The radiation hydrodynamics equations, which are given explicitly in the
co-moving frame of reference, are discretized numerically based on a Lagrangian
description and an adaptive mesh. They are solved implicitly, where the
advection scheme used is second-order total variation diminishing. Details can be obtained in
\citet{Liebendoerfer:etal:2002} and \citet{Liebendoerfer:etal:2004}. For the
current study, we include 104 radial adaptive mass zones. Resolution is adapted
according to high gradients in density, velocity, and entropy. Furthermore, the
outer boundary condition is set by applying a Schwarzschild metric at the outermost mass zone, assuming vacuum outside. In addition, neutrino transport is
discretized in 6 angular bins ($\cos\theta\in\{-1,...,+1\}$) and 20 energy bins
($E=3,...,300$~MeV).

In \citet{Liebendoerfer:etal:2004}, the authors compare the model with
S.~Bruenn's multi-group flux limited diffusion neutrino transport approximation.
A similar comparison with the Boltzmann transport used by the Garching group has
been published in \citet{Liebendoerfer:etal:2005}. These two studies show a good
qualitative agreement of the spherically symmetric models used by the different
groups.
\subsection{Use of the EOS}
\label{use}
The EOS in supernova simulations is part of the nuclear physics input, next to
weak interactions in the neutrino transport. There are two intrinsically
different regimes. In NSE, the destruction and
production of nuclei is in thermal and chemical equilibrium regarding strong and
electromagnetic interactions. The conditions for NSE are achieved typically
above a temperature of $T\sim0.5$~MeV.\footnote{Throughout the article we use
natural units, i.e., $\hbar=c=$k$_B=1$, for quantities where it is appropriate.}
For temperatures below $T= 0.44$~MeV, where NSE cannot be applied, we use an
ideal Si-gas. For the simulations with the HS and STOS EOS in the non-NSE regime
we calculate the baryon EOS as an ideal Maxwell--Boltzmann gas of silicon with
the measured nuclear binding energy from~\citet{Audi:etal:2003}. For LS this is not necessary, because the LS routines
used in the supernova model also provide the EOS of the silicon gas. The LS routines also handle intrinsically the transition to the NSE regime at temperatures of $T>0.44~$~MeV. 
For the HS and STOS EOS the baryon component in NSE is given in tabular form. It is a good approximation
to treat electrons (and positrons) as a rigid uniform background. Thus they can
be added as non-interacting ideal Fermi--Dirac gases based on
\citet{TimmesArnett:1999}. The blackbody contribution of photons is also taken
into account. 

The baryon EOS tables are stored for given temperature $T$, baryon
number density $n_B$, and total proton fraction $Y_p=n_p^{\rm tot}/n_B$. Because of charge neutrality, in the absence of
muons the number density of electrons $n_e$ is given by $n_e=Y_p n_B$ which
connects the baryon and the electron EOS. Accordingly, $Y_p$ is equivalent to
the electron fraction $Y_e=n_e/n_B=Y_p$. All particle densities and fractions used in the article correspond to net densities, i.e., the difference between particle and antiparticle densities, e.g., $n_e=n_{e^-}-n_{e^+}$, $Y_e=Y_{e^-}-Y_{e^+}$. 
In astrophysics, very often instead of
the baryon number density $n_B$ a baryon mass density $\rho$ is used. This is
completely equivalent, if $\rho$ is defined as $\rho = n_B m$, with an arbitrary
but constant mass $m$. We will also use such a ``baryon mass density'' with the
usual convention $\rho = n_B m_u$, with the atomic mass unit $m_u =
931.49432$~MeV. However, we want to stress that $\rho$ defined in such a way is
not the rest-mass density, because, e.g., the dominant particle species in NSE
depends on the thermodynamic state of the system. In a relativistic
description of the EOS with interactions there is no conservation of rest mass, but only of the total energy
and baryon number. Analogously to the baryon mass density one also defines a
baryonic mass $M= N_B m_u$ which is just a redefinition of the total baryon
number $N_B$ of the investigated system. 

The HS EOS contains the full distribution of all available nuclei, as will be
discussed in more detail in the next subsection. However, regarding weak
processes we only consider an average heavy nucleus and an average light
nucleus. We separate the distribution of all nuclei into light nuclei and heavy
nuclei by the charge number six, i.e. carbon:
\begin{eqnarray}
X_a&=&\sum_{A\geq2,Z\leq 5} A\ n\az/ n_B \; ,
\label{massfracs1}
\\
X_A&=&\sum_{A\geq2,Z\geq 6} A\ n\az/ n_B  \; ,
\label{massfracs2}
\end{eqnarray}
with $n\az$ denoting the number density of a nucleus with mass number $A$ and
charge number $Z$, so that the sum of the mass fractions of neutrons $X_n$,
protons $X_p$, light nuclei $X_a$ and heavy nuclei $X_A$ is unity. The average
mass and charge of the light and heavy nuclei are:
\begin{eqnarray}
\langle a \rangle &=& \sum_{A\geq2,Z\leq 5}A\ n\az \  /  \sum_{A\geq2,Z\leq
5}n\az, \\
\langle z \rangle &=& \sum_{A\geq2,Z\leq 5} Z\ n\az \  /  \sum_{A\geq2,Z\leq
5}n\az,
\end{eqnarray}
\begin{eqnarray}
\langle A \rangle &=& \sum_{A\geq2,Z\geq 6} A\ n\az \ /  \sum_{A\geq2,Z\geq
6}n\az, \\
\label{mass1}
\langle Z \rangle &=& \sum_{A\geq2,Z\geq 6} Z\ n\az \  /  \sum_{A\geq2,Z\geq
6}n\az \; .
\label{mass2}
\end{eqnarray}

We further simplify the neutrino reactions with light nuclei by treating all of
them as alpha particles. Note that for the alpha particles, and thus for all
light nuclei, in the present investigation only neutral--current reactions are
considered, as listed in Table~\ref{tab:nu-reactions}. Charged-current and other
break-up reactions of the weakly bound light nuclei may also be important in
supernova simulations, see, e.g., \citet{Nakamura:etal:2009} for neutrino deuteron
reactions. However, in this article we first want to investigate their possible
appearance and their contribution to the EOS. 
\subsection{Model of the HS EOS}
\label{sec_nse}
The HS model consists of an ensemble of nucleons and nuclei in NSE, whereas
interactions of the nucleons and excluded volume corrections for the nuclei are
implemented. Note that in the present version of the HS EOS tables, the
formation of nuclei is neglected for $T>20$~MeV for simplicity, i.e., matter is
assumed to be uniform. At such high temperatures there is only a limited
density region slightly below nuclear matter saturation density $n_B^0$, where
nuclei appear at all. Still it would be better to avoid such a hard switch in
the EOS, which we plan to do in future releases of the EOS tables. In the
following, we give a brief summary of the HS model, all details can be found in
\citet{HempelSchaffnerBielich:2010}. 

For the unbound interacting nucleons (neutrons and protons) an RMF model is applied. Its Lagrangian is based on the exchange of
the isoscalar--scalar $\sigma$, the isoscalar--vector $\omega$ and the isovector--vector 
$\rho$ mesons between nucleons. In comparison with more sophisticated
theoretical approaches, based on nucleon--nucleon scattering data, it has been
recognized that nonlinear $\sigma$ and $\omega$ self-coupling terms are
necessary to achieve a reasonable description of the EOS at high densities and
the properties of nuclei at the same time, see, e.g., \citet{SugaharaToki:1994}.
An alternative approach, which is not used in the present investigation, is RMF
models with density-dependent couplings, see, e.g., \citet{typel2005}. The free
parameters of the Lagrangians, the meson masses and
their coupling strengths, have to be determined by fits to experimental data. 

We apply three different RMF parameterizations: TM1 by
\citet{SugaharaToki:1994}, TMA by \citet{Toki:etal:1995}, and FSUgold by
\citet{ToddRutelPiekarewicz:2005}. The Lagrangians of TM1 and TMA have the same
form and include nonlinear terms of the $\sigma$ and $\omega$ mesons. TM1 was
developed together with TM2, which were fitted to binding energies and charge
radii of light (TM2) and heavy nuclei (TM1). TMA is based on an interpolation of
these two parameter sets. The coupling parameters $g_i$ of the set TMA are
chosen to be mass-number dependent of the form $g_i=a_i+b_i/A^{0.4}$, with $a_i$
and $b_i$ being constants, to have a good description of nuclei over the entire
range of mass numbers. For uniform nuclear matter the couplings become constants
and are given by $a_i$. TMA was also used in
\citet{HempelSchaffnerBielich:2010}, where the HS model was introduced. FSUgold
includes the coupling between the $\omega$ and the $\rho$ meson in addition.
This leads to a better description of nuclear collective modes, the EOS of
asymmetric nuclear matter and a different density dependence of the symmetry
energy \citep{Piekarewicz:2007} which is very low at large densities. The
coupling constants of FSUgold are fitted to binding energies and charge radii of
a selection of magic nuclei.

The HS EOS tables take into account the experimental data on nuclear masses from
\citet{Audi:etal:2003}. For the masses of the experimentally unknown nuclei
different theoretical nuclear structure calculations in form of nuclear mass
tables are used. The TMA interactions are combined with the mass table of
\citet{Geng:etal:2005}, which is also calculated with the RMF model TMA. Thus
all nuclear interactions are consistent. This mass table lists 6969 even--even,
even--odd, and odd--odd nuclei, extending from $^{16}_{~8}$O to $^{331}_{100 }$Fm
from slightly above the proton to slightly below the neutron drip line. The
nuclear binding energies are calculated under consideration of axial
deformations and the pairing is included with a BCS-type $\delta$-force. For the
parameterization TM1 we do not have a suitable mass table at hand, thus we cannot
avoid the minor inconsistency to also use the table of \citet{Geng:etal:2005},
which is based on the TMA parameterization. For FSUgold we take a mass table
calculated by X.~Roca-Maza, which was previously applied for the outer crust of neutron stars 
\citep{RocaMazaPiekarewicz:2008}. This table contains 1512 even--even nuclei,
from the proton to the neutron drip, with $14\leq A \leq 348$ and $8\leq Z\leq
100$. Odd nuclei are not included in this table. The nuclei were calculated only
with spherical symmetry and the pairing is introduced through a BCS approach
with constant matrix elements. The constant matrix element for neutrons has been
fitted to reproduce the experimental binding in the tin isotopic chain and the
constant matrix element for protons to the experimental binding in the $N=82$
isotonic chain.

To describe nuclei in the supernova environment, we not only need binding
energies, but have to account for medium and temperature effects. For the
screening of the Coulomb field of nuclei in the uniform background of
electrons we use the most basic expression: for each nucleus we assume a
spherical Wigner--Seitz (WS) cell at zero temperature. More elaborated approaches
for the Coulomb energy of a multi-component plasma at finite temperature can,
e.g., be found in
\citet{NadyozhinYudin:2005}, \citet{Potekhin:etal:2009}, and \citet{PotekhinChabrier:2010}. However,
we leave this for future studies as the Coulomb energy becomes only important at
low temperatures so that the simplest expression is sufficient for our purposes.

Finite temperature leads to the population of excited states of nuclei. Here,
we use the temperature dependent degeneracy function of \citet{FaiRandrup:1982}.
It is the same analytic expression as in the original reference of the HS model
\citep[][]{HempelSchaffnerBielich:2010} and used previously by \citet{Ishizuka:etal:2003}, but now we consider only excitation
energies below the binding energy of the corresponding nucleus, in order to
represent that the excited states still have to be bound (see, e.g.,
\citet{Roepke:1984}). We note that the inclusion of excited states up to
infinite energies had only a minor influence on the composition but would lead
to an unphysically large contribution of the excited states to the energy
density and entropy at very high temperatures.

We describe nuclear matter as a chemical mixture of the different nuclear
species and nucleons. As we distinguish between nuclei and the surrounding
interacting nucleons we still have to specify how the system is composed of the
different particles. Our thermodynamic model is built on two main assumptions:
first, we assume for unbound nucleons that they are not allowed to be situated
inside of nuclei, whereas nuclei are described as uniform hard spheres at
saturation density $n_B^0$. Second, for nuclei (with mass number $A\geq 2$) we
assume that they must not overlap with any other baryon in the system (nuclei or
unbound nucleons). Thus, we take the volume which is available for the nucleons
to be the part of the total volume of the system which is not excluded by
nuclei. This is described by the filling factor of the nucleons
\begin{equation}
\xi=1-\sum \az A~n\az/n_B^0 
\end{equation}
(here and in the following, we refer to $A\geq 2$). The free volume in which a
nucleus can move is the total volume minus the volume filled by nuclei and
nucleons. This is incorporated via the free volume fraction
\begin{eqnarray}
\kappa&=&1-n_B/n_B^0 \; ,
\label{nse_eq_kappa}
\end{eqnarray}
with the total baryon number density $n_B$, which includes contributions from
unbound neutrons and protons:
\begin{eqnarray}
n_B&=& n_n+n_p+\sum \az A~n\az \; .
\end{eqnarray}

Based on these two main assumptions, the EOS is derived in a consistent way,
using the non-relativistic Maxwell--Boltzmann description for nuclei and the full
Fermi--Dirac integrals for nucleons (solved with the routines from
\citet{Aparicio:1998} and \citet{gong:2001}). We obtain modifications of all
thermodynamic quantities due to the excluded volume, in agreement with the
results of \citet{yudin11} who studied excluded-volume schemes in general. Here
we give the thermodynamic potential, the free energy density $f$, as an
example: 
\begin{eqnarray}
f
=&&
\sum \az f\az^0(T,n\az) + \sum \az f\az^{\rm Coul} 
\nonumber \\
&&-
T\sum \az n\az \mathrm{ln}(\kappa) + \xi f_{\rm RMF}^0(T,n_n/ \xi,n_p/ \xi)\; .
\label{eq_f}
\end{eqnarray}
The first term in Eq.~(\ref{eq_f}) is the summed ideal gas expression of
nuclei. The Coulomb free energy of nuclei appears in addition. The third term is the direct contribution from the excluded volume.
Because of this term, as long as nuclei are present, the free energy density
goes to infinity when approaching saturation density, because the free volume of
nuclei goes to zero, $\kappa \rightarrow 0$. Thus, nuclei will always disappear
before saturation density is reached. The RMF contribution of the nucleons,
$f^0_{\rm RMF}$, is weighted with their filling factor $\xi$, as the free energy is
an extensive quantity. If nuclei are absent, $ \xi = 1$, and we get the
unmodified RMF description, as it should be. The excluded volume correction for
the nuclei represents a hard-core repulsion of the nuclei at high densities
close to saturation density. Instead the modification of the free energy of the
unbound nucleons is purely geometric and just describes that the nucleons fill
only a fraction of the total volume. In this sense, the two aforementioned model
assumptions for the excluded volume are essential, as they lead to the desired
limiting behavior of the EOS.
\subsection{EOS Characteristics and Constraints}
\label{subsec_char}
Table \ref{nse_table_rmf} lists some characteristic saturation properties of
uniform bulk nuclear matter for the three different RMF parameterizations.
We also include results for the LS EOS with the
incompressibilities of $K=180$~and 220~MeV. The quantities shown in Table
\ref{nse_table_rmf} correspond to the coefficients of the following power-series
expansion of the binding energy per baryon at $T=0$ around the saturation
point: 
\begin{eqnarray}
E(x,\beta)&=& -E_0 + \frac{1}{18}Kx^2 +\frac{1}{162}K' x^3 + \cdots \nonumber \\
 && +\beta^2\left(J+\frac{1}{3} L x + \cdots\right) + \cdots \; ,
\end{eqnarray}
with $x=n_B/n_B^0-1$ denoting the relative deviation from the saturation
density, and the asymmetry parameter $\beta$ which is given by the total proton
fraction: $\beta= 1-2 Y_p$. Next we discuss the listed nuclear matter properties
briefly and compare with experimental constraints. The saturation point itself,
i.e., the saturation density $n_B^0$ and the corresponding binding energy per
nucleon $E_{0}$, can be determined from the analysis of nuclear binding
energies. All EOS models give reasonable values for $n_B^0$ and $E_{0}$. The
nuclear incompressibility $K$ can be studied experimentally, e.g., by exciting
the isoscalar giant monopole resonance (ISGMR), leading to values of $K=240 \pm
10$~MeV \citep{piekarewicz10}. FSUgold and LS~(220) agree roughly with this
experimental constraint, TM1 is still reasonable, but the value of TMA is too
large and the value of LS~(180) too low. However, it is perceived in the
literature that the extraction of $K$ from ISGMR data is not unambiguous. The
experiments probe nuclear matter slightly below saturation density and
furthermore the deduced results depend on the nuclear interactions (e.g., the
behavior of the symmetry energy) which are taken for the data analysis (see, e.g., \citet{shlomo06,piekarewicz06,sharma09}). One also still has
difficulties to explain all available ISGMR data consistently with a single
theoretical model \citep{piekarewicz10}. The skewness coefficient $K'$ is the
third-order term for the expansion of the energy per nucleon around $n_B^0$ and
is not constrained directly by experiments so far. Still there exist
some estimates: based on a correlation study of Skyrme--Hartree--Fock models,
\citet{chen12} deduces $K' = - 355 \pm 95$~MeV. Only TM1 and LS (220) lie within
this region.

\begin{table}[t]
\caption{Properties at Saturation Density of the Different EOS Under
Investigation}
\centering
\begin{tabular}{c c c c c c c}
\hline
\hline
&
$n_B^0$ 
&
$E_{0}$ 
&
$K$ 
&
$K'$ 
&
$J$ 
&
$L$ 
\\
EOS & {[fm$^{-3}$]} & [MeV] & [MeV] & [MeV] &[MeV] & [MeV] \\
\hline
TM1 & 0.145 & 16.3 & 281 & --285 & 36.9 & 111 \\
TMA & 0.147 & 16.0 & 318 & --572 & 30.7 & 90 \\
FSUgold &  0.148 & 16.3 & 230 & --524 & 32.6 & 60 \\
LS (180) & 0.155 & 16.0 & 180 & --451 & 28.6 & 74 \\
LS (220) & 0.155 & 16.0 & 220 & --411 & 28.6 & 74 \\
\hline
\hline
\end{tabular}
\flushleft
{\bf Note.}
Definition of the quantities given in the text.
\label{nse_table_rmf}
\end{table}

Typical values for the symmetry energy coefficient $J$ range from 28 to 34~MeV,
whereas most studies give mean values around 32~MeV, see, e.g., \citet{tsang09}.
Thus the symmetry energy of TM1 does not comply with the experimental results,
but all other EOS do. We remark that the values of $J$ shown for LS, which we
calculated with the numerical routines which are available online, agree only
approximately with the value given in the original work by
\citet{LattimerSwesty:1991} of $J =29.3$~MeV. The slope of the symmetry energy,
$L$,  shows an important correlation with the neutron skin thickness of heavy
nuclei \citep{rocamaza11} which is currently being measured by the PREX
experiment at JLAB. Interestingly, $L$ is also correlated with the radii of
neutron stars \citep{horowitz01}. There are various experimental probes for $L$,
leading to constraints ranging from $40$ to $110$~MeV with a common center
around 75~MeV, see the comprehensive compilations of \citet{carbone10} and
\citet{tsang09}. In some nuclear models a large value of $J$ is compensated by a
large value of $L$ in the fitting procedure of the free parameters, to achieve a
similar symmetry energy slightly below saturation density. This is also seen for
TM1, which gives too large values of $J$ and $L$. The other investigated EOS give
reasonable results for $L$. 

We want to emphasize the following, simple point which will reappear several
times during the later discussion of our results: single parameters do not
characterize the global behavior of the EOS. To give one example, one could
think that TM1 is softer than TMA because of the lower incompressibility.
However, the slight change of the saturation point and less negative higher
order terms such as the skewness coefficient in TM1 lead to the unexpected
result that the pressure of symmetric nuclear matter in TM1 is always larger
than in TMA, i.e., TM1 is stiffer. Furthermore, for supernovae and neutron stars
the symmetry energy plays an important role. The extremely large
symmetry energy of TM1 stiffens its neutron star EOS additionally. Thus for
TM1 one obtains a maximum mass of cold neutron stars, which is significantly
larger than the one of TMA, as we will see next.

\begin{figure}[t]
\centering
\includegraphics[width=0.9\columnwidth, clip=true]{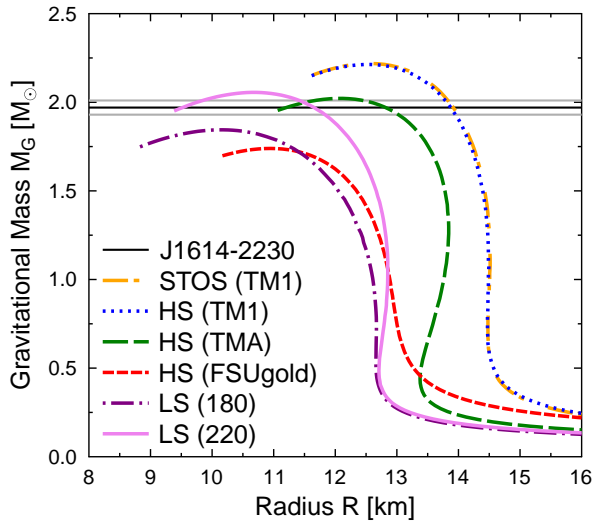}
\caption{Mass--radius relation of cold neutron stars for the EOS under
investigation. The gray and black horizontal lines show the measurement of the
mass of the pulsar PSR J1614-2230 with its 1$\sigma$ error
\citep{Demorest:etal:2010}.}
\label{fig:mr}
\end{figure}

In 2010, the gravitational mass of the millisecond pulsar PSR J1614-2230 was
determined to be $M_G = 1.97 \pm 0.04\:$M$_\odot$ by measuring the Shapiro delay
\citep{Demorest:etal:2010}. It represents the largest robust maximum mass
measurement at present and thus the most important constraint for the
mass--radius relations of neutron stars. In Fig.~\ref{fig:mr}, we show this
constraint with its 1$\sigma$ error together with the mass--radius relations of
the EOS which are investigated in this article. The curves have been
calculated with the tables which are used in the simulations presented below by
imposing beta-equilibrium without neutrinos at practically zero temperature,
i.e., $T=0.1$~MeV, and solving the Tolman--Oppenheimer--Volkoff equations. The
values of the corresponding maximum masses of neutron stars are 2.219~M$_\odot$,
2.213~M$_\odot$, 2.022~M$_\odot$, 1.739~M$_\odot$, 1.844~M$_\odot$, and
2.056~M$_\odot$ for STOS~(TM1), HS~(TM1), HS~(TMA), HS~(FSUgold), LS~(180), and
LS~(220), respectively. The mass--radius curves of STOS~(TM1) and HS (TM1) are
almost identical, because they are built with the same RMF parameterization TM1.
This good agreement shows that there are only small differences in the
pressure--energy density relation between the two EOS models, despite the
different description of non-uniform nuclear matter. FSUgold and
LS~(180) fail to fulfill the constraint of PSR J1614-2230. Despite this, we
include FSUgold in the present study, to explore the implications of a very soft
RMF EOS and because of its good description of low-density nuclear matter. We
also keep LS~(180), because it is a standard reference for core-collapse
supernova simulations. The stiff TM1 parameterization lies well above the
maximum mass constraint, whereas TMA and LS~(220) are only slightly above.

\citet{steiner2010} used observational data of masses and radii of seven neutron
stars with well-determined distances in a Bayesian framework to determine
nuclear matter parameters and the neutron star EOS. The found values for the
saturation properties of nuclear matter agree remarkably well with the
aforementioned experimental constraints. Furthermore, the authors deduce a
mass--radius relation which is based on the analysis of the observational data
and find relatively small neutron star radii around 11 to 12 km for $M_G=
1.4$~M$_\odot$, indicating a soft behavior of the neutron star EOS around
saturation density. Only LS~(180) is compatible with such small
radii, and FSUgold and LS~(220) come close to it. The radii of TMA and TM1 seem
to be much too large.

We conclude that TMA has satisfying nuclear matter properties, apart from the
too large value of the incompressibility. For TM1 the incompressibility is
barely acceptable, but the symmetry energy and its slope are too large. On the
other hand, these two EOS are not compatible with small neutron star radii
found by \citet{steiner2010}. FSUgold has very nice nuclear matter properties
and a lot of success in nuclear structure, but unfortunately it gives a too low
maximum mass. Also LS~(180) with an incompressibility of 180~MeV is actually
ruled out by the observation of PSR J1614-2230, and furthermore its
incompressibility and symmetry energy are too low. LS~(220) gives
satisfactory results. It would be good if other new EOS tables became available
in the future which comply better with all the experimental and observational
results. The emphasis of the present study is a different one. We want to
explore a broad range of possible EOS, including extreme cases, to better
understand the impact of the EOS in core-collapse supernovae.

\begin{figure}
\centering
\includegraphics[width=\columnwidth, clip=true]{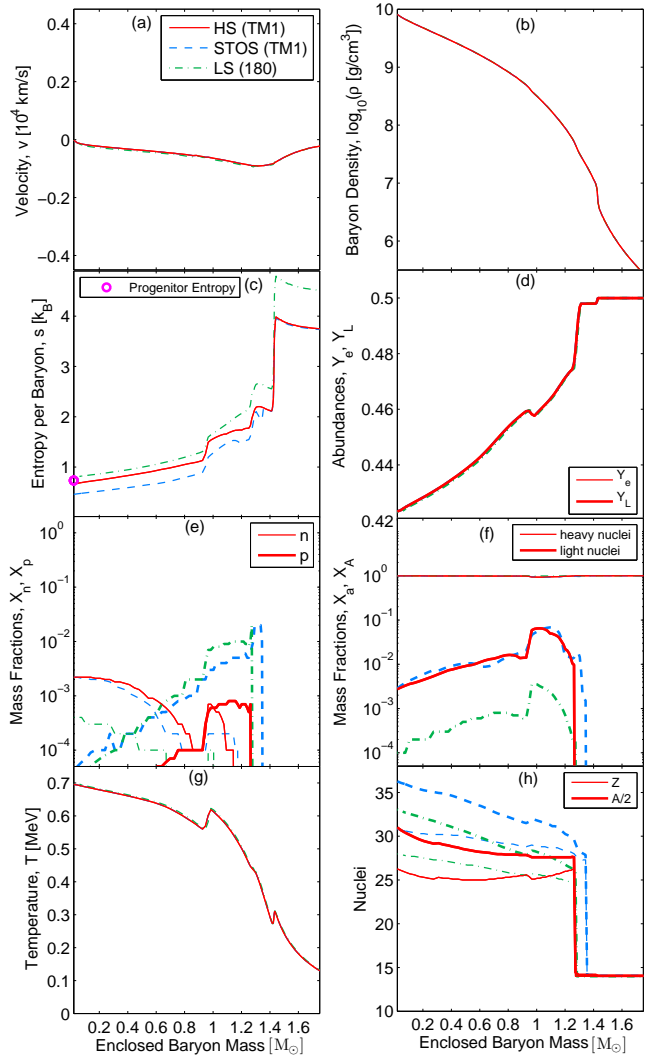}
\caption{Radial profiles of selected hydrodynamic quantities with respect to the
enclosed baryon mass for the 15~M$_\odot$ progenitor model from
\citet{WoosleyWeaver:1995} at the progenitor stage, comparing the different EOS
of HS~(TM1) (red solid lines), STOS~(TM1) (blue dashed lines) and LS~(180)
(green dash-dotted lines). In panel (d), $Y_e$ is the electron fraction, and
$Y_L$ the electron lepton fraction. ``light nuclei'' in panel (f) corresponds to
the mass fraction of alpha particles for STOS and LS, and the mass fraction
$X_a$ of light nuclei with $Z\leq 5$ for HS. ``heavy nuclei'' is the mass fraction
of the representative heavy nucleus for STOS and LS, and the mass fraction $X_A$ of
all nuclei with $Z\geq 6$ in HS. Panel (h) shows the charge and mass number of
the representative heavy nucleus for STOS and LS, respectively, the average
charge and mass number of the distribution of heavy nuclei for HS. See the text for further details.}
\label{fig:fullstate-progenitor}
\end{figure}
\section{Results: general features of the HS EOS}
\label{sec_15}
In the following subsections, we will compare the HS EOS with the
parameterization TM1 (HS~(TM1)), introduced in \S~2, with the standard EOS from
\citet{LattimerSwesty:1991} with an incompressibility of 180~MeV (LS~(180)) as
well as H.~\citet{Shen:etal:1998} (STOS~(TM1)). Note that STOS is also based on
the TM1 parameterization. We examine the collapse, bounce and early post-bounce
phases at the example of the 15~M$_\odot$ progenitor model from
\citet{WoosleyWeaver:1995}. In this section, we do not consider the LS~(220) EOS
or any of the other RMF parameterizations of the HS EOS. Thus in the following
discussion sometimes we omit the specification ``(180)'' for the ``LS~(180)''
EOS. The same holds for the ``HS~(TM1)'' and ``STOS~(TM1)'' EOS, where we
sometimes use only ``HS'' or ``STOS'' for simplicity.

\begin{figure*}
\centering
\includegraphics[width=0.95\textwidth, clip=true]{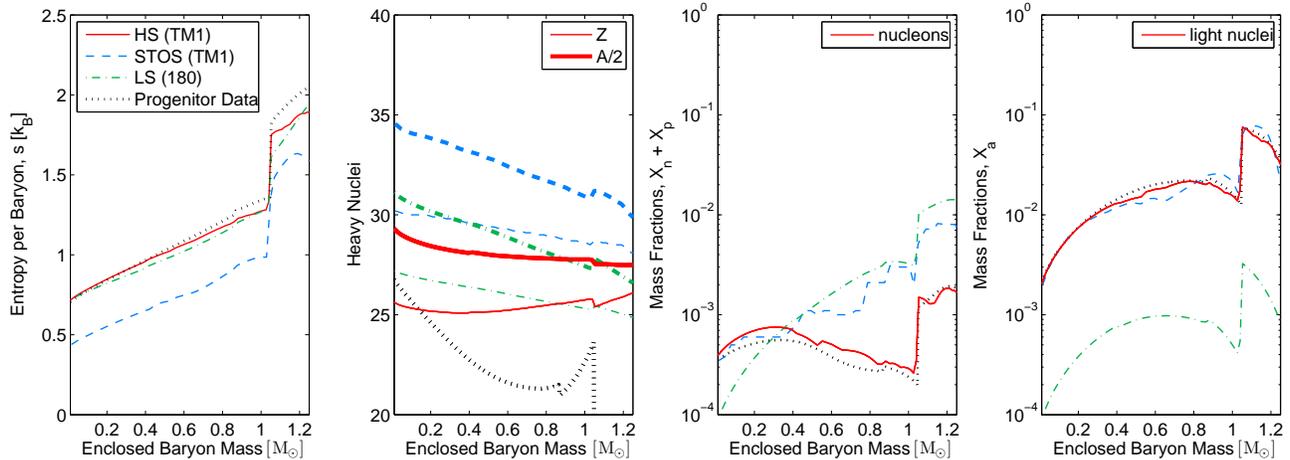}
\caption{Radial profiles of the entropy per baryon and the nuclear composition for the
15~M$_\odot$ progenitor model from \citet{Woosley:etal:2002} at the progenitor
stage, with the same notation as in Fig.~\ref{fig:fullstate-progenitor}. In
addition, we also compare with the progenitor data (black dotted lines) given by
the stellar evolution calculation. Note that only the average mass number, but
not the charge number of heavy nuclei, was available in the progenitor data and
is depicted in the figure.}
\label{fig:progenitor}
\end{figure*}
\subsection{Progenitor Stage}
The differences between the EOS can already been identified at the progenitor
stage, illustrated via radial profiles in Fig.~\ref{fig:fullstate-progenitor}
for the 15~M$_{\odot}$ progenitor of \citet{WoosleyWeaver:1995}. We note that we
take the temperature, density, and electron fraction from the progenitor as the
initial configuration for the core-collapse simulation, that is why these
quantities are the same for all three EOS. For the part of the contracting progenitor star
which is shown in Fig.~\ref{fig:fullstate-progenitor}, the baryon density ranges
from $10^{10}$ to $10^{5}$~g/cm$^3$ and the entropy per baryon from 0.5 to
5~k$_B$, corresponding to temperatures of 0.7--0.1~MeV. Matter at the center
is already slightly asymmetric, whereas the electron fraction ranges from $Y_e
\sim 0.42$ at the very center to 0.5 at the lowest densities shown in the
figure. Note that the electron lepton fraction $Y_L=Y_{\nu_e}+Y_e$, with the net electron-flavor neutrino fraction $Y_{\nu_e}$, is on top of
the electron fraction, as the abundance of electron neutrinos (and
all other neutrinos, too) is still negligible. One sees that the composition is
dominated by heavy nuclei, only at a region around $M_B \sim 1$~M$_{\odot}$
alpha particles contribute with up to 10~\% to the mass fraction.

Next we compare the state of the progenitor described by the three EOS in
detail. We note that the HS EOS should give the most accurate description of the
progenitor stage, as it includes shell effects and experimentally measured
binding energies. Medium modifications of the properties of nuclei are
negligible at such low densities and temperatures encountered here. The fraction
of light nuclei $X_a$ with $Z\leq 5$ in the HS EOS is almost identical to the
alpha-particle fraction of STOS, showing that the additional light nuclei are
negligible at the progenitor stage. LS gives a reduction of the alpha-particle
fraction of almost one order of magnitude, which can be attributed to a
well-known error in the alpha-particle binding energy in the LS routines
\citep{swesty2005}. In the reference simulations which we show here, we did not
correct for this error. Apart from this difference, the dependence of the
alpha-particle abundance on density is very similar in the three EOS. 

The fraction of unbound nucleons is much more model dependent, due to the
interplay between unbound nucleons and the formation of heavy nuclei. Regarding
heavy nuclei in the inner layers, there is a clear tendency of STOS to give
the largest mass numbers, followed by LS and HS with the smallest heavy nuclei.
As expected from stellar nucleosynthesis, iron-group nuclei are found in HS. In
contrast, STOS gives nuclei around $Z \sim 30$. Compared to the composition of
the progenitor (not shown in Fig.~\ref{fig:fullstate-progenitor}, because only
limited information was available), all EOS show some significant differences,
because NSE has actually not been reached completely in the progenitor
calculation, even though temperatures are above 0.5~MeV. For example, there
is still up to 20~\% of carbon in the very center. Also thermodynamic quantities
are affected from the fact that the progenitor is not in full NSE. For example,
the central entropy per baryon of the progenitor model (purple circle in
Fig.~\ref{fig:fullstate-progenitor}(c)) is slightly larger than in HS, because the
average mass number of heavy nuclei is lower than the value obtained from NSE.
LS gives a similar but a bit larger entropy than HS. In contrast, the entropy
of STOS is significantly lower, which we will explain in the discussion of the
15~M$_{\odot}$ progenitor of \citet{Woosley:etal:2002} shown in
Fig.~\ref{fig:progenitor}.

We discuss this progenitor too, because the available data include entropy
profiles and more detailed information about the nuclear composition.
Furthermore, this progenitor is closer to NSE than the one of
\citet{WoosleyWeaver:1995} shown before. Regarding the composition, similar
trends of the three EOS as in Fig.~\ref{fig:fullstate-progenitor} are observed.
In comparison with the original progenitor data, there is a perfect agreement
with HS for the mass fractions of the different particles. However, the average
mass number of heavy nuclei from the progenitor calculation is still below the
NSE value from HS. HS and LS reproduce the entropy profile of the progenitor
very well, but the entropy per baryon of STOS is almost 0.3~k$_B$ lower. For the
shown conditions, the entropy is mainly given by heavy nuclei. The entropy of
heavy nuclei can be split in two contributions: internal excitations and kinetic
entropy from the thermal movement of nuclei. All three EOS take into account internal excitations of nuclei, but describe it in
different ways. Furthermore, only HS has a distribution of nuclei, whereas STOS
and LS are based on the SNA. However, the largely reduced entropy of STOS is
mainly due to the absence of the kinetic entropy of nuclei. The nuclear kinetic
entropy per baryon is on the order of 0.2~k$_B$ in LS and HS, which explains the
reduction of the entropy in STOS seen in Figs.~\ref{fig:fullstate-progenitor}
and \ref{fig:progenitor}. This means that STOS assumes always a lattice of
nuclei, even if a crystal would actually form only at lower temperatures. Also
the energy density, the pressure, and the chemical potentials of STOS lack in the
kinetic contribution of nuclei in the ideal gas regime at low densities and
moderate temperatures. This is a typical problem of SNA models. LS resolves this
problem by adding an explicit kinetic contribution of heavy nuclei to the free
energy. 
\subsection{Core Collapse and Bounce}
\begin{figure}
\centering
\includegraphics[width=\columnwidth, clip=true]{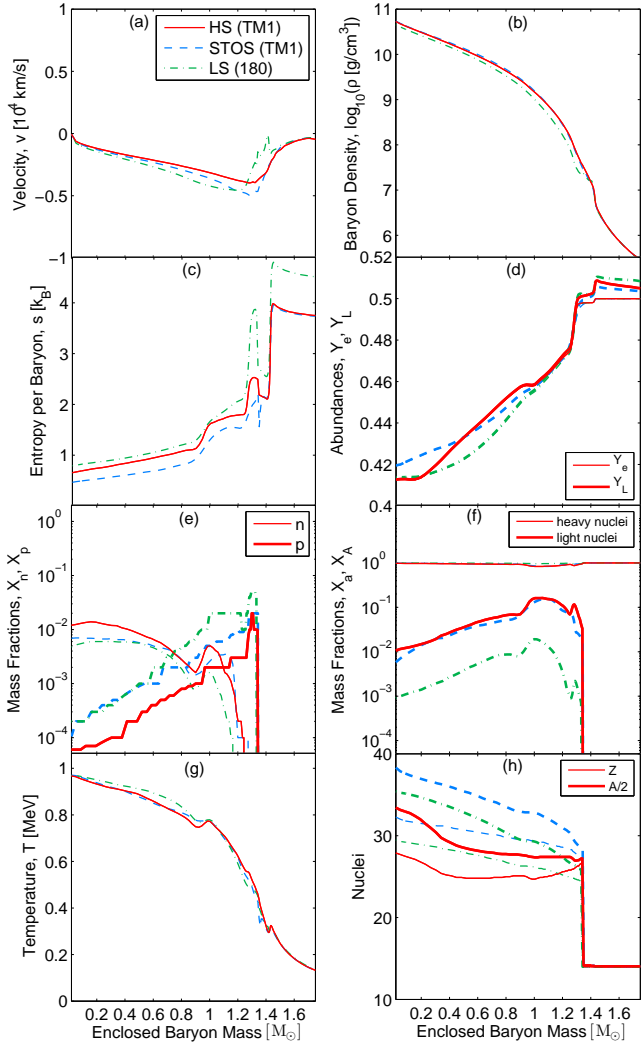}
\caption{Radial profiles of selected hydrodynamic quantities with respect to the
enclosed baryon mass during the core-collapse phase for the 15~M$_\odot$
progenitor model from \citet{WoosleyWeaver:1995}, when the central temperature
$T_\text{central} \simeq 1$~MeV. Shown are the HS~(TM1) EOS at $t_{\rm pb}=-42.9$~ms
(red solid lines), the STOS~(TM1) EOS at $t_{\rm pb}=-42.1$~ms (blue dashed lines),
and the LS~(180) EOS at $t_{\rm pb}=-39.1$~ms (green dash-dotted lines). The
notation is the same as in Fig.~\ref{fig:fullstate-progenitor}.}
\label{fig:fullstate-collapse1}
\end{figure}

To follow the evolution of the core-collapse phase, in
Fig.~\ref{fig:fullstate-collapse1}, we show radial profiles of selected
quantities for the three EOS under investigation (LS~(180), STOS~(TM1),
HS~(TM1)), when the central temperatures reach $\sim1$~MeV. The selected states
correspond to slightly different times in the simulation around 40~ms before
bounce, but represent similar stages of the evolution. For the collapse phase
until the stage shown in Fig.~\ref{fig:fullstate-collapse1}, the evolution of
the inner core proceeded homologously. No shock has formed and any possible
change of the entropy per baryon is given only by weak reactions. As long as the
densities are low, neutrinos emitted can leave without interactions. As
explained in \citet{Bruenn:1985} and \citet{Liebendoerfer:2005}, during the
early stage of collapse the entropy generation by electron captures is
roughly balanced by the entropy carried away by neutrinos. Here, we also find
that the entropy profiles remain almost unchanged. However, a large difference
of the entropy per baryon developed at $\sim1.4$~M$_\odot$ between the three
EOS, which relates to the different transitions from non-NSE to NSE. The huge
difference between LS and the other two EOS is caused by the different non-NSE
treatment for LS mentioned in \S~\ref{use}. It is to some extent artificial
and can be safely ignored. Due to the ongoing compression, the matter in the
entire collapsing core is heated slightly, which in turn increases the fraction
of alpha particles and nucleons by almost one order of magnitude. By comparing
the average mass and charge numbers of heavy nuclei, one sees that some electron
captures have taken place, and that the mass numbers have been increased. These
processes and the evolution occur in an overall similar manner for the three
different EOS. The differences which were observed for the progenitor in
Fig.~\ref{fig:fullstate-progenitor}, i.e., larger nuclei in STOS and LS,
different nucleon fractions, and a too low entropy in STOS, can still be
identified.

However, some quantities which were identical at the progenitor stage developed
differently. Let us focus on the comparison of HS~(TM1) and STOS~(TM1) which are
built from the same nuclear interactions. For example, the electron fraction
shows an interesting evolution. Up to 0.4~M$_{\odot}$, HS leads to a faster
deleptonization than STOS. We explain this by the used prescription of electron
captures on heavy nuclei by \citet{Bruenn:1985}, where the rate decreases with
the excess neutrons of a nucleus. Even though $\langle Z \rangle/\langle A
\rangle$ is roughly the same in HS and STOS, because nuclei in STOS are
significantly larger, the number of excess neutrons is larger, leading to
smaller electron-capture rates. We leave the question whether this effect would
persist in the more elaborated descriptions of electron capture on heavy nuclei
of \citet{Hix:etal:2003} and \citet{Langanke:etal:2003} for future studies.
Between $0.4~\text{M}_{\odot} < M_B < 1~\text{M}_{\odot}$, the deleptonization
was slower in HS than in STOS. This is the domain which is dominated by electron
captures on free protons. Because the free-proton fraction is larger in STOS
than in HS (see Fig.~\ref{fig:fullstate-collapse1}(e)) electron captures in STOS
are faster. Because the electron contribution to the EOS is dominating in this
region, the smaller electron fraction of STOS still has a small effect on hydrodynamic
variables like the infall velocity or the baryon density, which are slightly
increased in STOS.

\begin{figure}[h]
\centering
\includegraphics[width=\columnwidth, clip=true]{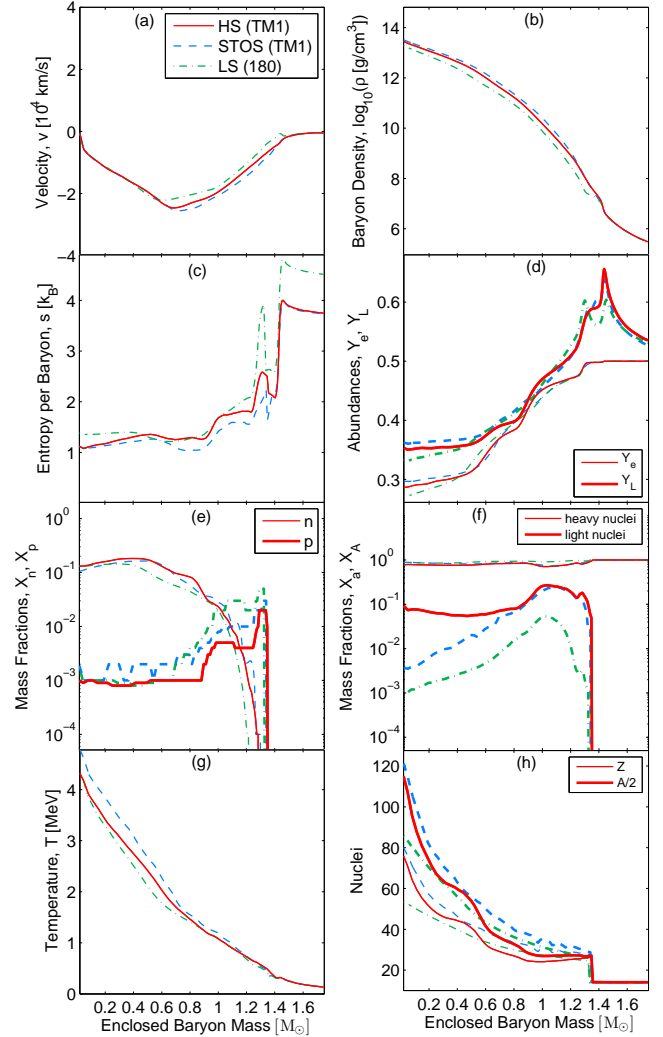}
\caption{Same as Fig.~\ref{fig:fullstate-collapse1} but at $T_\text{central}
\simeq 5$~MeV. The states correspond to $t_{\rm pb}=-0.91$~ms for HS~(TM1),
$t_{\rm pb}=-0.84$~ms for STOS~(TM1), and $t_{\rm pb}=-1.09$~ms for LS~(180).}
\label{fig:fullstate-collapse2}
\end{figure}

Figure~\ref{fig:fullstate-collapse2} shows the evolution of the collapse when
the central temperatures reach roughly 5~MeV corresponding to $\sim$ 0.1~ms
before bounce. By comparing the fraction of light nuclei $X_a$ of HS with the
alpha-particle fraction of STOS in Figs.~\ref{fig:fullstate-collapse1} and
\ref{fig:fullstate-collapse2} (thick lines in graphs~(f)) one realizes that
other light nuclei than alpha particles have been formed with HS in the center during
these two stages of the collapsing star. The total fraction of light nuclei in
the center is almost 10\% in Fig.~\ref{fig:fullstate-collapse2}, whereas the
alpha particle fraction has decreased below 1\%. If weak processes with the
additional light nuclei were taken into account, in principle this could also
modify the neutrino transport and deleptonization. However, the fraction of
light nuclei other than alpha particles starts to rise only shortly before the
stage shown in Fig.~\ref{fig:fullstate-collapse2}. For example, the fraction of
$^3$H gets larger than the alpha-particle fraction around 2~ms before bounce. At
this stage, neutrinos are already trapped in the center. Thus we expect that the
additional light nuclei have only little influence on the deleptonization.

At the stage of Fig.~\ref{fig:fullstate-collapse2} there is a significant number
of trapped neutrinos. At the center of the star the electron and lepton
fractions in HS are still lower than in STOS. The change of the electron
fraction at the center during the stages of Fig.~\ref{fig:fullstate-collapse1}
and Fig.~\ref{fig:fullstate-collapse2}, i.e., the number of electron captures is
very similar for the two EOS: $\Delta Y_e^\text{HS}=-0.1261$ and $\Delta
Y_e^\text{STOS}=-0.1222$. Also the change of the central lepton fraction was
similar for the two EOS, they give a similar amount of deleptonization in the
center. In the outer layers the differences in the lepton fraction profiles are
less pronounced and it appears that STOS and LS are closer to each other than
STOS and HS. Interestingly, the electron and lepton fraction profiles show a
more step-like change in HS. These steps coincide with larger changes of the
average mass and charge number of heavy nuclei which are caused by neutron magic
shell effects, as will be further illustrated below. 

In the profiles of the entropy per baryon an increase of $\sim 0.5$~k$_B$ is
observed in the center. The entropy creation occurs between the two stages of
Figs.~\ref{fig:fullstate-collapse1} and \ref{fig:fullstate-collapse2}, before
neutrinos become trapped and reach weak equilibrium. This happens at the stage
where the entropy loss by neutrino transport is smaller than the entropy gain
from local weak reactions. If we compare the entropy profiles of HS and STOS we
see that the difference of the entropy per baryon for $M_B >
0.6~\text{M}_{\odot}$ is the same as it was at the progenitor stage. Contrary,
in Fig.~\ref{fig:fullstate-collapse2} at the center, HS and STOS have equal
entropies per baryon, thus there was a larger entropy production in STOS of
$\Delta s^\text{STOS} \sim 0.65$~k$_B$ than in HS of $\Delta s^\text{HS} \sim 0.47$~k$_B$.
Where does this additional entropy come from? Based on the work of
\citet{Bruenn:1985}, \citet{Liebendoerfer:2005} showed that the entropy
production can be well approximated by
\begin{eqnarray}
T ds &=&-dY_e(\Delta \mu - E_{\nu}^{\text{esc}}) \; , \\
\Delta \mu &=& \mu_e+\mu_p-\mu_n\; ,
\end{eqnarray}
where $E_{\nu}^{\text{esc}}$ denotes the typical energy of escaping neutrinos
which is on the order of 10~MeV, and $\Delta \mu>0$ is the average chemical
energy which is liberated by an electron capture. $\mu_e$, $\mu_p$, and $\mu_n$
are the relativistic chemical potentials of electrons, protons, and neutrons,
i.e., they include the corresponding rest masses. Note that neutrinos are emitted
at a lower energy than they are produced, because they thermalize by inelastic
scattering before escape. As the change of the electron fraction, the electron
neutrino luminosities and the temperature evolution are very similar in HS and
STOS, we find that the difference of the change of the entropy per baryon,
$\Delta s^\text{STOS}-\Delta s^\text{HS} \sim 0.2$~k$_B$, has to originate from a
difference $\Delta \mu^\text{STOS}-\Delta \mu^\text{HS}$ on the order of several MeV. This
means that electron captures in STOS give a larger release of entropy than in
HS. Indeed, by carefully comparing the chemical potentials of HS and STOS (see
\citet{HempelSchaffnerBielich:2010}) one finds that the non-relativistic proton
chemical potentials (i.e., without the proton rest mass) for the conditions
encountered here are slightly lower in HS than in STOS. This is due to the
different description of non-uniform matter and nuclei. Also light nuclei other
than alpha particles, which form during the two stages of the collapsing star
shown in Figs.~\ref{fig:fullstate-collapse1} and \ref{fig:fullstate-collapse2}
(graphs~(f)), give a small modification of the chemical potentials in HS. On the
other hand, STOS has in general larger electron fractions in the center, so that
$\mu_e$ will be larger, leading to an increase of $\Delta \mu^\text{STOS}$ compared
to $\Delta \mu^\text{HS}$. This means that electron captures are more energetic,
because matter in STOS is still more symmetric than in HS. 

Furthermore, the nucleon rest masses employed in the EOS have an interesting
effect on the entropy production as we will discuss now. The EOS model of STOS
does not include the neutron-to-proton rest-mass difference, but only a common
nucleon mass of 938~MeV. Contrary, HS is built on the measured neutron and
proton masses. For the conditions encountered here, the nucleon rest masses
contribute directly to the energy and free energy densities and the relativistic
chemical potentials. Thus, the rest-mass difference is included in  $\Delta
\mu^\text{HS}$ but not in $\Delta \mu^\text{STOS}$, which alone changes $\Delta \mu$
already by $\Delta m \sim 1.2935$~MeV. We conclude, that the appearance of light
nuclei, the faster deleptonization in the early phases of the collapse and
realistic nucleon masses lead to less entropy gain for electron captures in HS. 

We want to emphasize that in the electron/positron capture rates based on
\citet{Bruenn:1985} which determine $dY_e$, the nucleon rest-mass difference
$\Delta m$ is taken into account, independent of the EOS. This is done by using
the chemical potentials without rest mass and explicitly adding $\Delta m$ to
the $Q$-value. Thus, in both simulations the energy of the produced neutrinos and
the rate of deleptonization is determined correctly taking into account the
rest-mass difference of the nucleons. The missing rest-mass difference in STOS
leads only to a slight overestimation of the entropy production, as discussed
before. This also explains why the change of the electron fraction is so similar
for the two EOS, despite the different entropy production. In this context we
also want to remark that the nucleon rest masses have only a negligible effect
on the equilibrium (NSE) composition, as long as the same rest masses are used
for all nucleons, i.e., for free nucleons and nucleons bound in nuclei.

Let us turn to the temperature evolution. In Fig.~\ref{fig:fullstate-collapse1},
the central temperature was equal for STOS and HS. Now the temperature of
STOS has increased to larger values than in HS which is in agreement with the
larger entropy increase in STOS. However, one also has to note that the
densities reached in HS are slightly smaller than in STOS, so there is less
compression heating. In this context we have to remark that minor differences
can also be due to the different data sampling, because profiles which are shown
in the same figure can correspond to slightly different evolutionary states.
However, we checked that the differences which we see and which we discuss here
are systematic and not caused by the different data sampling.

\begin{figure}[t]
\centering
\includegraphics[width=\columnwidth, clip=true]{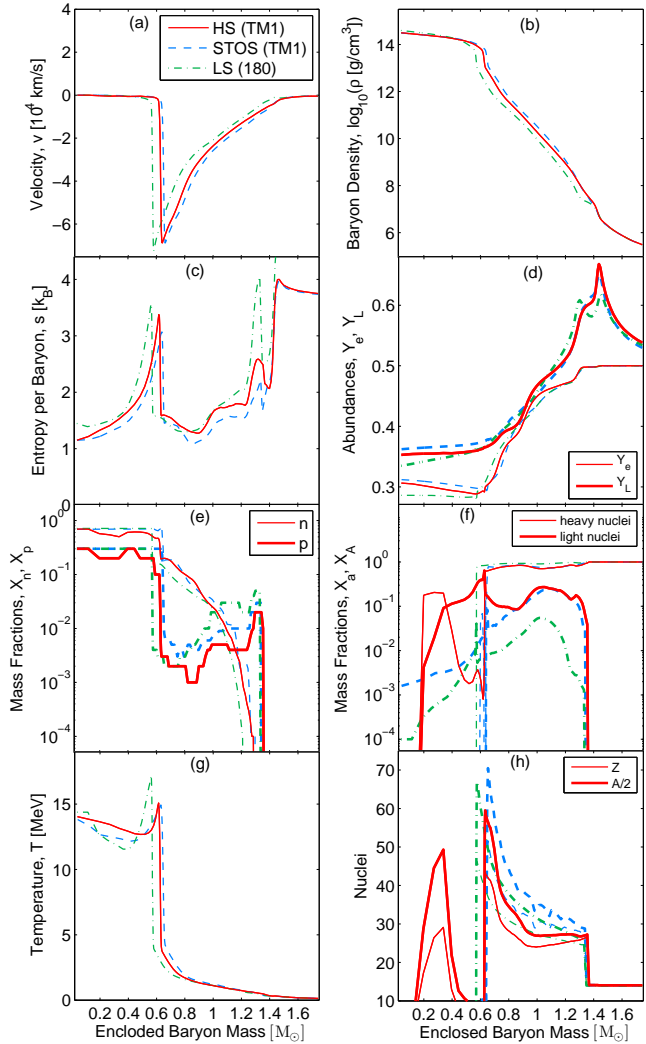}
\caption{Same as Fig.~\ref{fig:fullstate-collapse1} but at
bounce, defined as the moment when the maximum central density is obtained at
the end of the iron-core collapse.}
\label{fig:fullstate-bounce}
\end{figure}

Less than 1~ms later, the core bounce takes place, which is depicted
in Fig.~\ref{fig:fullstate-bounce}. It is defined as the moment when the maximum
central density is obtained at the end of the iron-core collapse. Regarding the
$Y_e$ and entropy profiles we can identify the same differences between HS and
STOS as described before. From Fig.~\ref{fig:fullstate-bounce} we can see that
the shock is located at a slightly lower mass in HS. It is well known that the
mass of the inner core is proportional to $Y_e^2$
\citep[see][]{GoldreichWeber:1980, Martinez:etal:2006}, which explains this
result to some extent. During the collapse evolution the temperature of HS was
similar to the one of STOS. Now at bounce we see that the core temperature of HS
has become larger than in STOS. It can also be attributed to the lower
electron fraction at the center. As electrons provide a large contribution to
the pressure at saturation density, the pressure in HS is lower, leading to more
compression and larger temperatures. We emphasize again that STOS and HS are
identical above saturation density. For the very soft non-relativistic LS, the
same differences to STOS appear but are even more pronounced: the shock forms at
smaller mass, the infall velocities are larger, and the central densities are
higher. The central conditions at bounce, comparing the three EOS, are listed in
Table~\ref{tab:bounce}, whereas $s$ denotes the entropy per baryon. We conclude
that even though HS and STOS are based on the same nucleon interactions, the
different description of non-uniform nuclear matter and nuclei leads to a
different evolution of the iron-core collapse. HS has a lower central electron
fraction, which results in a more compact configuration at bounce.

\begin{table}[t]
\caption{Selected Central Hydrodynamic Quantities at Bounce for the 15~M$_\odot$
Models}
\centering
\begin{tabular}{c c c c c}
\hline
\hline
EOS & $\rho$ & $s$ & $Y_e$ & $M$ \footnote{Baryon mass enclosed inside the
shock.}
\\
& $[10^{14}$ g/cm$^3 ]$ & [k$_B$] &  & $[$M$_\odot]$
\\
\hline
LS~(180) & 3.948 & 1.4463 & 0.2860 & 0.5799 \\
STOS~(TM1) & 3.061 & 1.1512 & 0.3120 & 0.6586 \\
HS~(TM1) & 3.135 & 1.1484 & 0.3063 & 0.6258 \\
\hline
\hline
\end{tabular}
\flushleft {\bf Note.} \label{tab:bounce} \end{table}

Regarding the composition there are two main differences between HS, and STOS
and LS: in the mass shells where the shock has ran through in HS there is a
significant fraction of more than 10\% of light nuclei present. The
alpha-particle fraction of LS and STOS is almost two orders of magnitude smaller
(see Fig.~\ref{fig:fullstate-bounce}(f)). Obviously, such a large abundance of
light nuclei influences the EOS. The second main difference in the composition
of the different EOS concerns heavy nuclei. At densities larger than $10^{14}$
g/cm$^3$ there is a small region where heavy nuclei appear in HS (see
Fig.~\ref{fig:fullstate-bounce}(h) between 0.2 and 0.4~M$_\odot$). This can be seen
as the beginning of the uniform nuclear matter phase and part of the transition
to uniform nuclear matter. Instead in STOS and LS no heavy nuclei form behind
the shock. One observes a different behavior of the transition to uniform
nuclear matter in HS in comparison to LS and STOS. This is also the density
regime, where the formation of the nuclear pasta phases is possible. In HS the
pasta phases are not taken into account, a Maxwell construction is used, and
furthermore the transition depends crucially on the implementation of the
excluded-volume effects. However, at these high densities neutrinos are
trapped. Hence, the heavy nuclei at high densities should have a negligible
effect on the neutrino transport especially since the heavy nuclei lie outside
the approximated Gamow--Teller window of \citet{Bruenn:1985}.
\subsection{Post-bounce Evolution}
\begin{figure}
\centering
\includegraphics[width=\columnwidth]{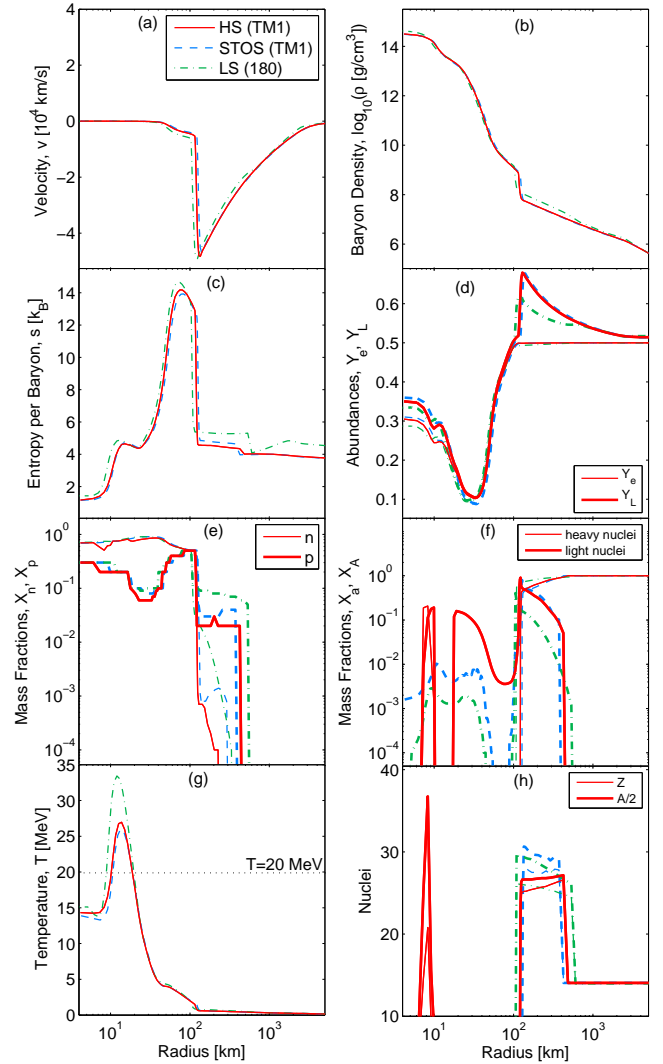}
\caption{Radial profiles of selected hydrodynamic quantities with respect to the
radius for the 15~M$_\odot$ progenitor model from \citet{WoosleyWeaver:1995} at
200~ms post bounce. The same notation as in Fig.~\ref{fig:fullstate-progenitor}
is used.}
\label{fig:fullstate-pb}
\end{figure}

Figure \ref{fig:fullstate-pb} shows the evolution for the three different EOS at
200~ms post bounce. The bounce shock stalls for all models and turns into the
standing accretion shock, which stands around 100~km at 200~ms post bounce (see
Fig.~\ref{fig:fullstate-pb}(a)). It separates the inner high-density and
high-temperature core from the outer material which is being accreted. Note the
entropy differences ahead of the shock between LS and STOS/HS, which are due to
the aforementioned different treatment of non-NSE. During the long-term
post-bounce evolution of several hundreds of milliseconds, the standing
accretion shocks contract in a similar fashion for all EOS under investigation
and hence the profiles become increasingly similar. The differences obtained at
bounce due to the different composition and different description of nuclei
remain in the post-bounce evolution. Matter is slightly more neutron-rich for HS
than for STOS, and HS leads to higher central densities and temperatures. In
comparison to LS, both HS and STOS are significantly stiffer. With LS the shock
contracts much faster and the central density and temperature are much higher,
as well as $Y_e$ is lower (see Fig.~\ref{fig:fullstate-pb}(a), (b), (c) and
(g)).

Also the composition still shows some similarities to the situation at bounce.
The infalling matter is composed of heavy nuclei, which are systematically
smaller for HS than in STOS and LS. The accreted matter is heated up and
dissociates partly into alpha particles. When matter reaches the accretion front
it encounters strong shock heating, so that nuclei get dissolved and almost only
free nucleons remain. Further inside the proto-neutron star the compression
becomes strong enough that the fraction of light nuclei increases again. In
general, alpha particles and light nuclei are favored in the innermost layers
not only because of the higher densities, but also because of the lower
entropies. In HS there is a small region slightly below saturation density where
even heavy nuclei appear, before they dissolve into homogeneous nuclear matter
at slightly higher densities. However, the fraction remains small, less than
$20~\%$ (see Fig.~\ref{fig:fullstate-pb}(f)). Heavy nuclei cannot be found
below the shock with STOS or LS at the shown time. But also with STOS heavy
nuclei are present inside the core with significant mass fractions but only from
shortly after bounce until roughly 50~ms post bounce, when they disappear again.
For STOS on the other hand, alpha particles extend down to the center of the
proto-neutron star in Fig.~\ref{fig:fullstate-pb}(f). In this model, due to the
different description of the excluded volume effects, alpha particles can also
exist slightly above saturation density. However, more sophisticated studies
show that the Mott transition actually occurs at lower densities
\citep{Typel:etal:2010}.

By looking at the electron fraction in more detail, one sees that the envelope
of the proto-neutron star below the standing bounce shock has a larger
electron fraction with HS than in STOS. This can be attributed to the larger
symmetry energy of HS in comparison to STOS, due to the inclusion of the
additional light nuclei besides alpha particles in the HS model. Light nuclei
exist as additional degrees of freedom for symmetric nuclear matter, which lower
the chemical potential of symmetric nuclear matter and therefore decrease the
$Q$-values of electron captures. As charged-current reactions on light nuclei
are not taken into account in the present investigation, this leads to less
electron captures than in STOS. We see in Fig.~\ref{fig:fullstate-pb}(f) that
the fraction of light nuclei in HS behind the shock is roughly two orders of magnitude
larger than in STOS (and LS). At $R\sim 20$~km the light nuclei fraction
increases beyond 10~\% in the NSE models, but jumps to zero between 10 and 20~km,
because in the HS EOS nuclei are neglected for $T>20$~MeV for simplicity.

\subsection{Nuclear Composition from Collapse to Post Bounce}
\begin{figure}
\centering
\includegraphics[width=0.9\columnwidth, clip=true]{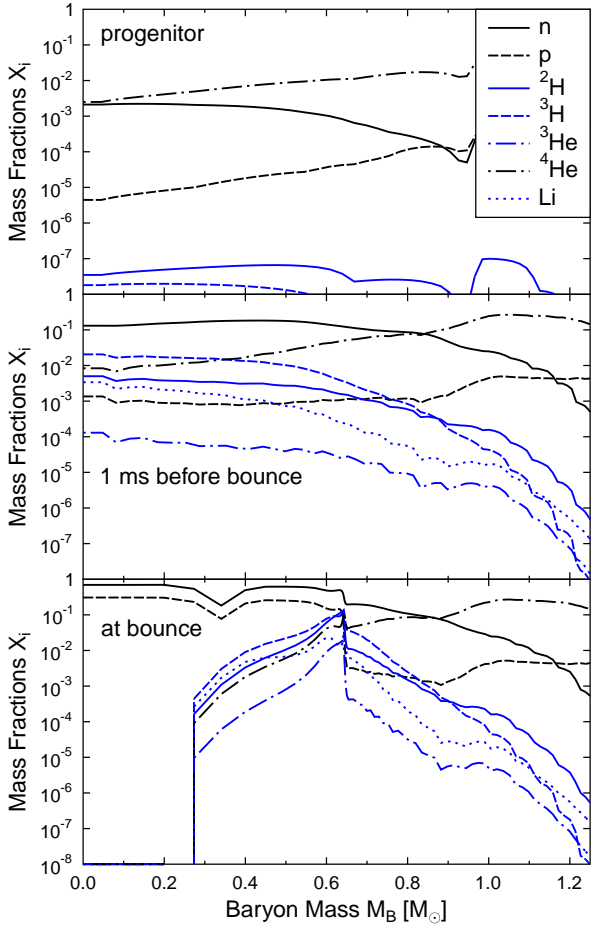}
\caption{Radial profiles of the mass fractions of the most important light
nuclei with $Z \leq 3$ in the HS (TM1) EOS for the core-collapse simulation of
the 15~M$_\odot$ from \cite{WoosleyWeaver:1995}, at the progenitor stage (top),
during the collapse (middle) and at bounce (bottom). ``Li'' shows the sum of the
mass fractions of all lithium isotopes.}
\label{fig:composition-light}
\end{figure}

Figure \ref{fig:composition-light} gives a more detailed view of the
contribution of light nuclei to the EOS. The mass fractions of the most
important light nuclei with $Z \leq 3$ are plotted for the simulation with HS
(TM1). The black lines depict the particles which are also included in LS and
STOS, and the blue lines depict light nuclei, which were usually not considered in
core-collapse supernova simulations until now. We want to emphasize that all
mass fractions which are shown in this article are a direct output from the EOS
applied in the simulations. In Fig.~\ref{fig:composition-light}, one sees that additional light nuclei apart from alpha
particles are negligible at the progenitor stage. This explains also the good agreement for the fraction
of light nuclei in Figs.~\ref{fig:fullstate-progenitor} and
\ref{fig:progenitor}. However, at later times during the core collapse where
temperature and density increase sufficiently, other light nuclei in addition to
alpha particles appear in non-negligible fractions on the order of a few
percent and they can become even more abundant than alpha particles. In general, the favored appearance of weakly bound light nuclei instead of alpha particles is driven by the increased
entropy of a system composed of smaller fragments, where binding energies play
only a minor role.  How does this work?

\begin{figure}[t]
\centering
\subfigure[$M_B$ = 0.6
M$_{\odot}$]{\includegraphics[width=0.8\columnwidth, clip=true]{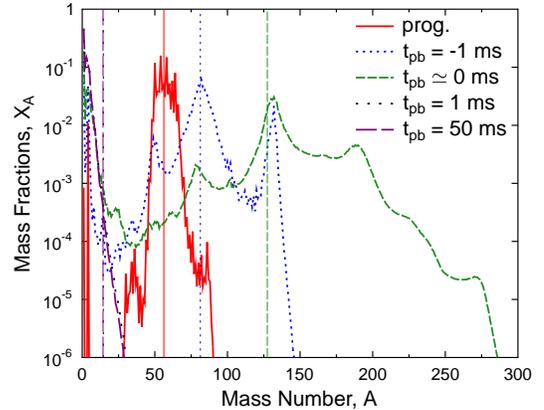}\label{fig:xa0.6a}}\\
\subfigure[$M_B$ = 0.8
M$_{\odot}$]{\includegraphics[width=0.8\columnwidth, clip=true]{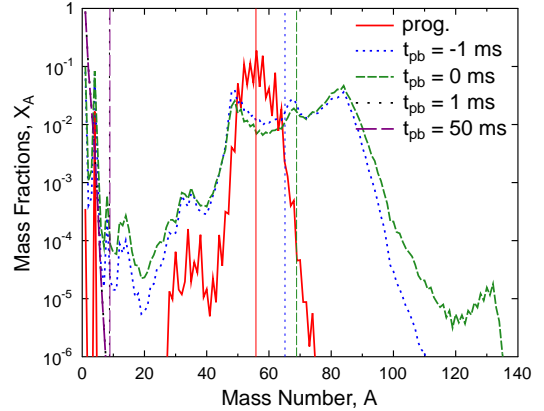}\label{fig:xa0.6b}}\\
\subfigure[$M_B$ = 1.0
M$_{\odot}$]{\includegraphics[width=0.8\columnwidth, clip=true]{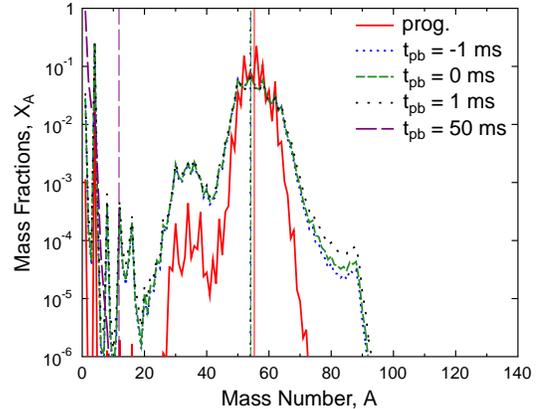}\label{fig:xa0.6c}}
\caption{Mass fraction distribution of the HS (TM1) EOS for three different
mass shells, and five different times, ranging from the progenitor to 50 ms post
bounce. The vertical lines show the corresponding average mass number of heavy
nuclei.}
\label{fig:xa0.6}
\end{figure}

Using Maxwell--Boltzmann distribution
functions for nuclei, schematically the number density of a nucleus of mass $A$ is 
\begin{equation}
 n_A \propto A^{3/2} \exp\left( A\frac{B_A+\mu}{T}\right) \; ,
\end{equation}
where $\mu$ is the
nucleon chemical potential without rest mass and $B_A>0$ the binding energy per nucleon of nucleus $A$. We remark that $B_A+\mu$ is
always negative in our calculations. Thus we do not see any indication for Bose--Einstein
condensation and the use of Maxwell--Boltzmann statistics is justified. For any given finite temperature, in the limiting case of vanishing density, $\mu$ will be so negative that $e^{A\mu / T}$ is completely dominating and the binding energies are negligible. Only free neutrons and protons will be present. This is an entropy effect, as a result of the minimization of the free energy. With increasing density the role of temperature decreases and at some point it is favorable to form nuclei, because the internal energy contribution to the free energy becomes dominating over the entropy contribution. Let us imagine that all nuclei had the same binding energy per nucleon $B$. In this scenario, the average mass number of the distribution of nuclei would increase continuously with density. Thus naturally there would be a certain density, where deuterons gave the main contribution to the composition, followed by alpha particles at larger densities. For sufficiently large temperatures $T\gtrsim B_A$, so that the differences in binding energies per nucleon of light nuclei can be seen as small corrections, this scenario gives a viable explanation for the large mass fractions of weakly bound light nuclei observed in our simulations. Contrary, for low temperatures $T<B_A$, the strong binding energy of the alpha particle is so dominant that other light nuclei will never give the main contribution to the composition. 

\begin{figure*}[t]
\subfigure[\ at the progenitor stage]
{\includegraphics[width=\columnwidth, clip=true]{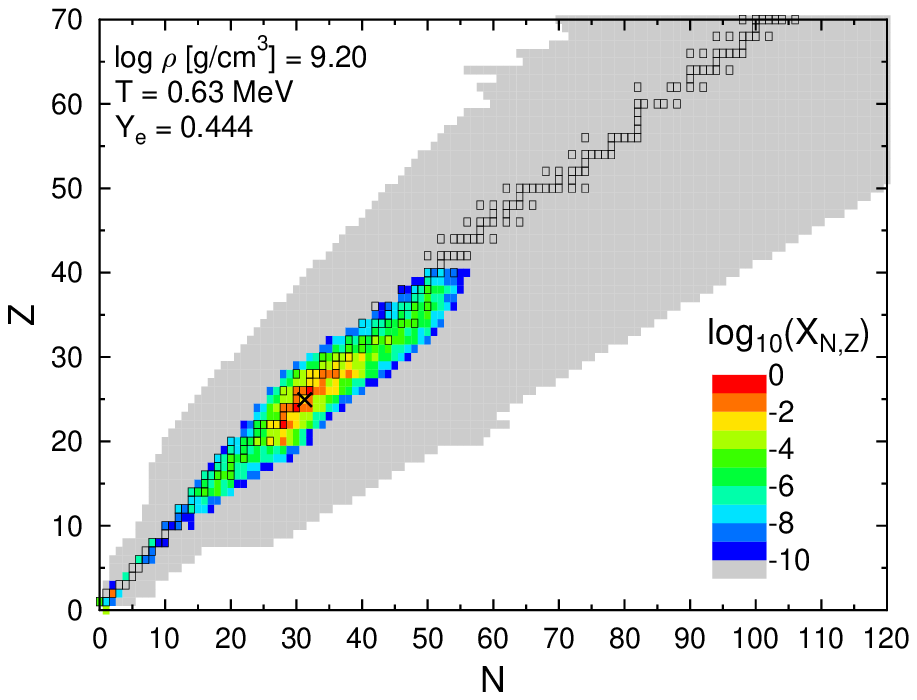}}
\hfill
\subfigure[\ 1 ms before bounce]
{\includegraphics[width=\columnwidth, clip=true]{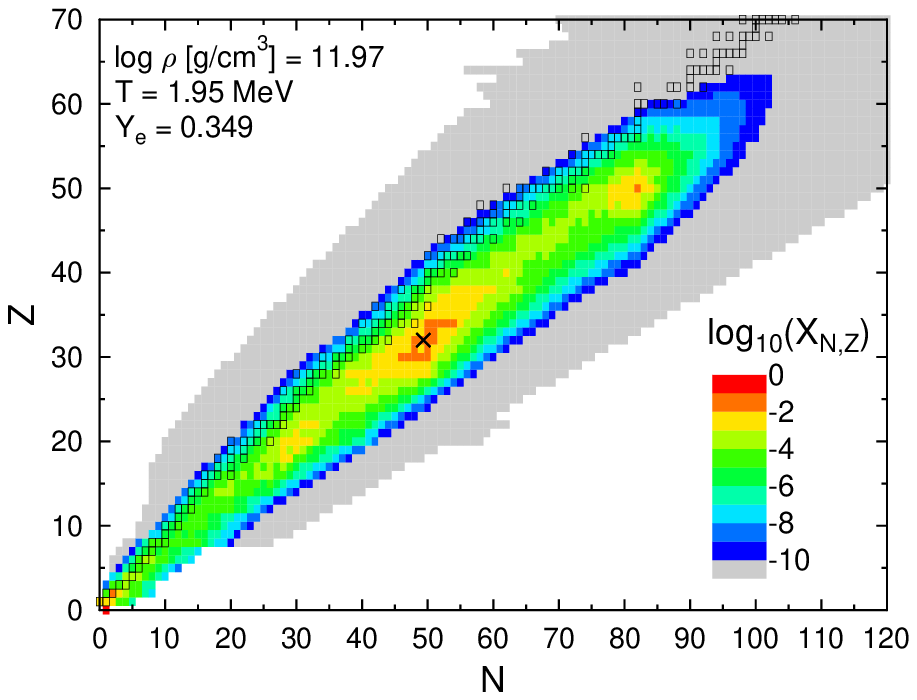}}
\\
\subfigure[\ directly before bounce]
{\includegraphics[width=\columnwidth, clip=true]{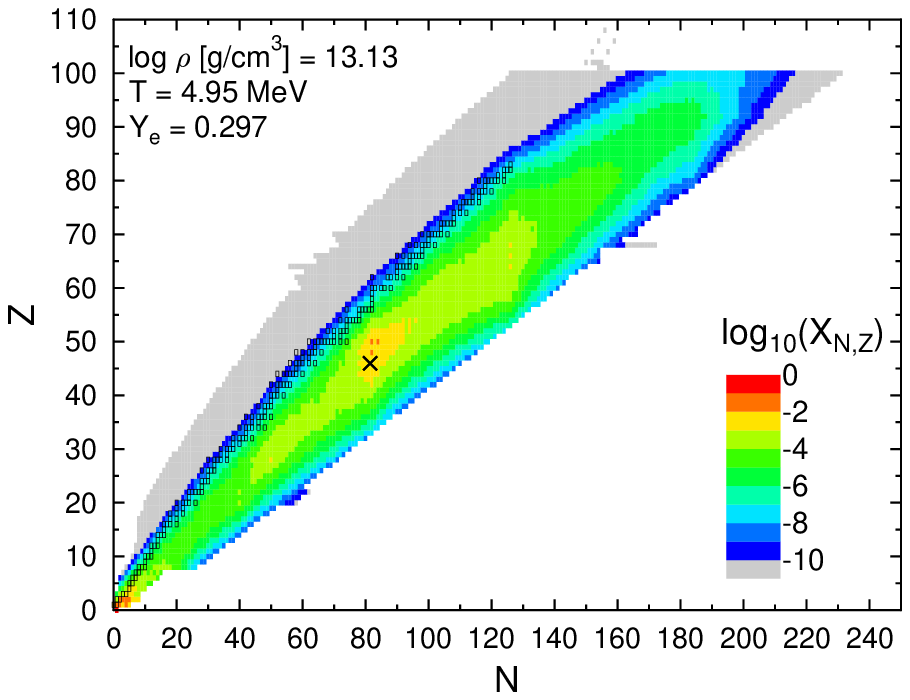}}
\hfill
\subfigure[\ 1 ms after bounce]
{\includegraphics[width=\columnwidth, clip=true]{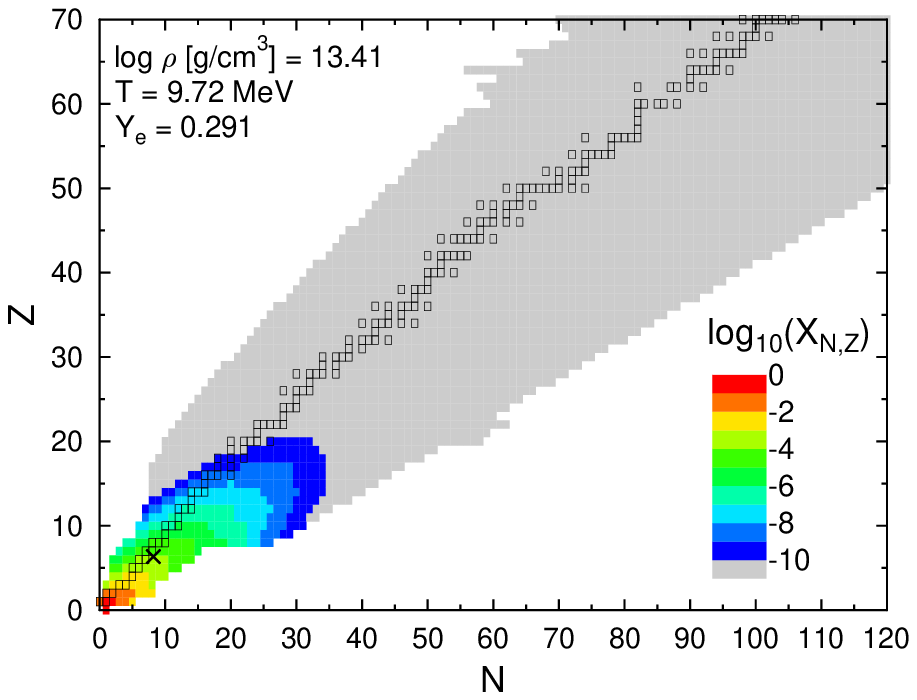}}
\caption{Distribution of the mass fractions of nuclei in the HS (TM1) EOS
for the mass shell of $M_B$ = 0.6 M$_{\odot}$ and four selected times, which are
also shown in Fig.~\ref{fig:xa0.6a}. Gray filled squares correspond to nuclei
which are included in the HS (TM1) EOS, but which have a mass fraction below
$10^{-10}$. The crosses mark the average of the distribution of heavy nuclei and
the black squares show stable nuclei as a reference.}
\label{fig:iso}
\end{figure*}

It is also instructive to realize that for any given finite temperature there is always a certain density where the mass fractions of deuterons and alphas are equal. This happens approximately at a chemical potential of $\mu \sim B_{d}-2B_\alpha \sim - 14$ MeV, where the difference of the total binding energies is compensated by the difference in total chemical potentials. At higher densities, the alpha fraction is larger, at lower densities the deuteron fraction. With increasing temperature this point moves to higher densities. Thus only for sufficiently large temperatures the fractions of deuterons and alphas will be comparable to the mass fraction of unbound nucleons. In conclusion, the significant appearance of weakly bound light nuclei is driven by an interplay of temperature and density effects.

Let us come back to the discussion of Fig.~\ref{fig:composition-light}. At 1~ms
before bounce, temperatures above 5~MeV are reached at the center (see
Fig.~\ref{fig:fullstate-collapse2}) which are sufficiently high so that light
nuclei can form at large densities. From Fig.~\ref{fig:composition-light} one
sees that tritium ($^3$H), alphas ($^4$He), deuterium ($^2$H), and lithium (Li)
appear in similar concentrations in the innermost layers of the collapsing core.
The largest fractions are reached for $^3$H because matter becomes increasingly
neutron rich during the core-collapse phase. We want to emphasize that the alpha
particle is less abundant than deuterium, tritium and the summed lithium
isotopes in the center. Only in the outer layers the composition is dominated by
alpha particles. 

At bounce (bottom of Fig.~\ref{fig:composition-light}) we see that the formation
of the shock enhances the appearance of unbound nucleons due to the increased
central densities and temperatures. At this stage, the matter behind the shock,
which is located at $\sim$ 0.6~M$_{\odot}$, is heated to roughly 14~MeV (see
Fig.~\ref{fig:fullstate-bounce}). Almost all heavy nuclei get dissociated, but
the fractions of light nuclei are increased. Note that the densities are very
large $\sim 10^{14}$ g/cm$^3$. At low densities there would be no light nuclei
for such large temperatures any more, but here we have a competition between
density and temperature effects, as described before. At the
very center of the collapsing stellar core, at densities above saturation
density, light nuclei disappear completely and uniform nucleon matter is
reached.

\begin{figure*}[t]
\centering
\subfigure[$\,\,\,\,$50~ms post bounce]{
\includegraphics[width=\columnwidth, clip=true]{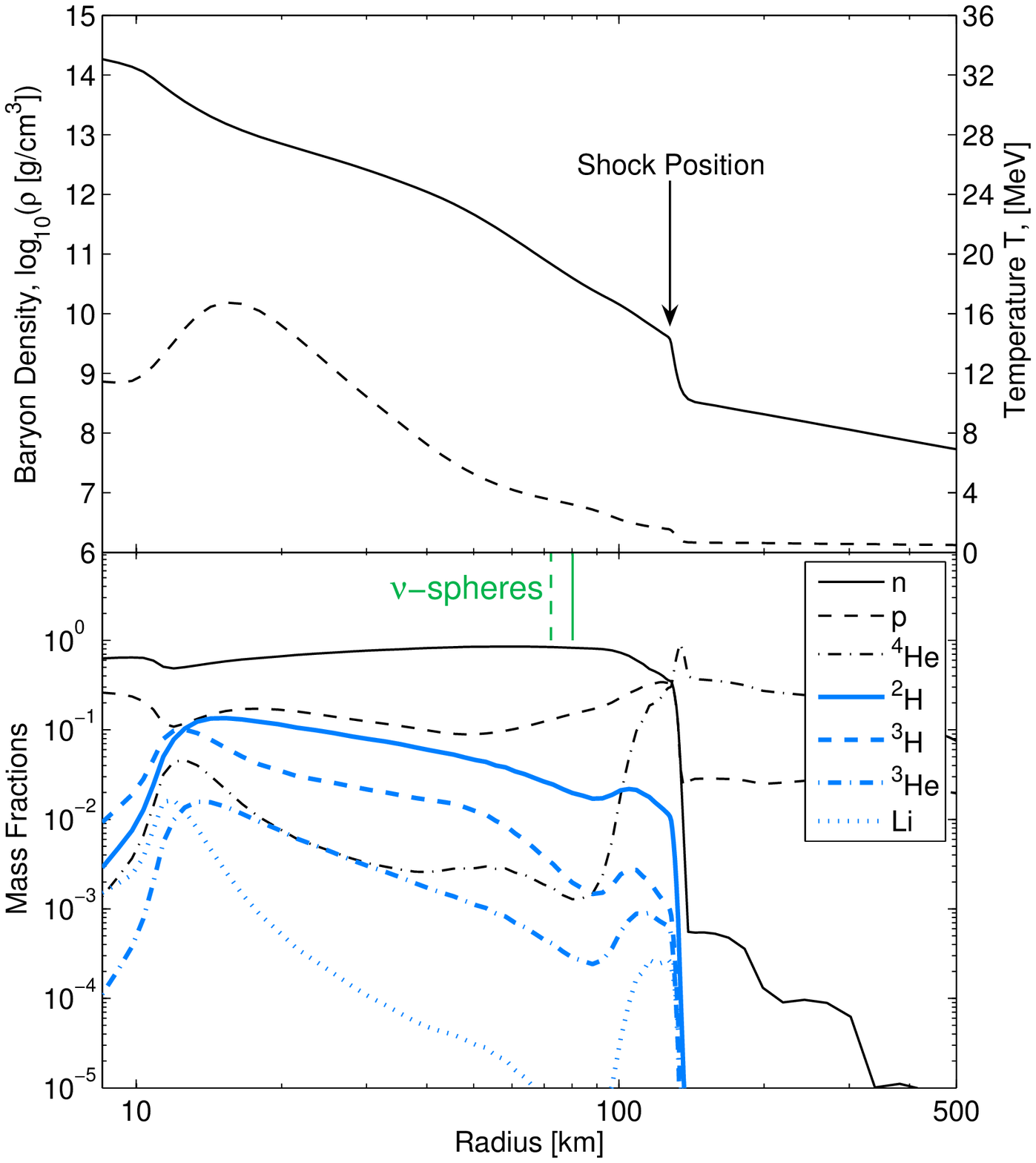}
\label{fig:comp-s15p-a}}
\hfill
\subfigure[$\,\,\,\,$200~ms post bounce]{
\includegraphics[width=\columnwidth, clip=true]{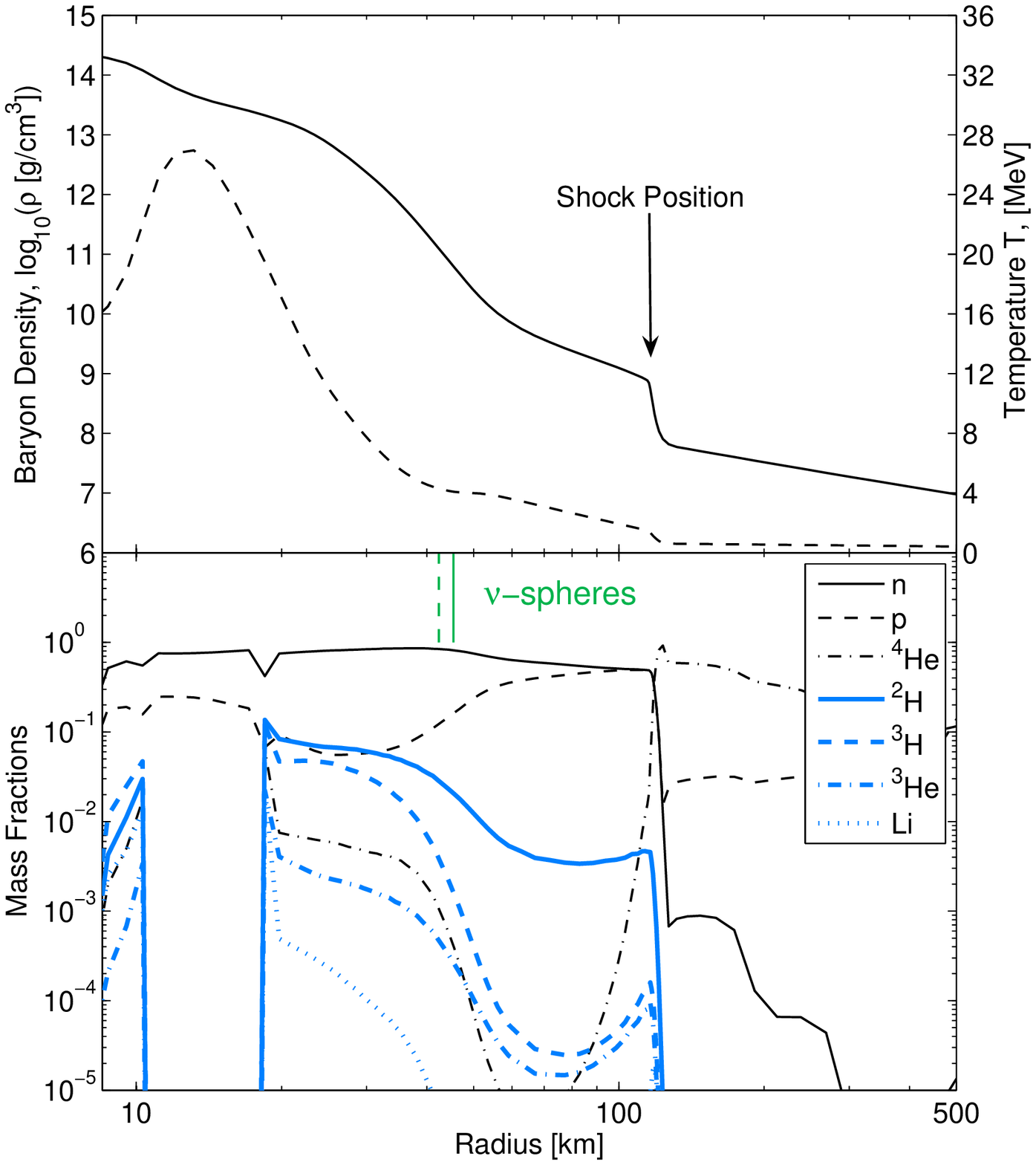}
\label{fig:comp-s15p-b}}
\caption{Radial profiles of the baryon density (solid lines) and
the temperature (dashed lines) in the top panels
and the corresponding nuclear composition regarding light nuclei (bottom panels)
at two selected post-bounce times for the 15~M$_\odot$ progenitor
model from \citet{WoosleyWeaver:1995} using the HS (TM1) EOS.
The green vertical lines denote the position of the neutrinospheres
for $\nu_e$ (solid lines) and for $\bar{\nu}_e$ (dashed lines).}
\end{figure*}

Regarding the contribution of heavy nuclei, with the HS EOS we can study the
impact of the commonly used SNA. Figure~\ref{fig:xa0.6}
shows the mass distributions of nuclei with the HS (TM1) EOS for the (baryonic) mass
shells of 0.6, 0.8, and 1.0~M$_{\odot}$ at different times. Fig.~\ref{fig:iso}
depicts the distributions in form of nuclear charts only for the 0.6~M$_{\odot}$
mass shell. We want to concentrate on this mass shell first. At the progenitor
stage (Fig.~\ref{fig:iso}(a) and red solid lines in Fig.~\ref{fig:xa0.6a}) we clearly
see that iron-group nuclei are most abundant, already mentioned in the
discussion of Fig.~\ref{fig:fullstate-progenitor}. The distribution is still
close to the valley of stability. During the ongoing contraction in the core
collapse, electron captures take place and the formation of heavier and more
neutron-rich nuclei is favored. In Fig.~\ref{fig:iso}(b), we see the importance
of closed neutron shells, i.e. nuclei with $N=50$ and $N=82$ are most abundant.
These nuclei correspond to the two peaks of the blue dotted lines in
Fig.~\ref{fig:xa0.6a}. Shortly before bounce, the distribution of heavy nuclei
becomes very broad and goes beyond $A=300$, which is nicely illustrated in
Fig.~\ref{fig:iso}(c). Note the different scale in this subfigure. Very exotic
nuclei are found and the distribution is far away from the valley of stability.
The most abundant nuclei are at the well-known neutron-magic number $N=82$, but
also other neutron-magic numbers (50 and 126) are strongly populated, visible by
the additional peaks and bumps in the distribution shown by the green dashed
lines in Fig.~\ref{fig:xa0.6a}. Obviously, such multiple peaks and/or bimodal
distributions cannot be captured by an average heavy nucleus, depicted by
vertical lines in Fig.~\ref{fig:xa0.6} and crosses in Fig.~\ref{fig:iso}.

Another interesting aspect of heavy nuclei is the largest mass numbers which
are reached in the simulations. \citet{furusawa11} pointed out that they obtain
nuclei with mass number up to 2000 in their EOS and even larger nuclei are
present in the STOS EOS. Superheavy nuclei with $A$ up to 3000 are also found in
the Hartree calculations by G.~\citet{shen2011a}. Contrary, in HS it is impossible to
obtain such superheavy nuclei, due to the use of nuclear mass tables based on experimental data and nuclear structure calculations. At the times shown in all the
previous plots the maximum average mass numbers are around 250, whereas all EOS
lie in a similar range. However, in Fig.~\ref{fig:iso}(c) one sees that the
distribution of heavy nuclei in HS reaches the border of the mass table. The
overall maximum mass numbers which we get during the entire simulation are $A
\sim 900$ for STOS, but only $A \sim 300$ for HS (TM1) and $A \sim 200$ for LS.
For all three EOS these overall maximum mass numbers are found at similar
conditions, at $\sim 0.5$~ms before bounce and $M_B  \sim 0.1$~M$_\odot$. From this point until bounce the mass numbers decrease, see Fig.~\ref{fig:fullstate-bounce}. In the
early post-bounce evolution of STOS there is a period where heavy nuclei appear
again in the core at densities around 10$^{14}$~g/cm$^3$, i.e., in the
transition region to uniform nuclear matter. At 50~ms post bounce these heavy
nuclei have disappeared again, therefore they cannot be seen in the previous
figures. This appearance of heavy nuclei is similar like the peak in the mass
fraction of heavy nuclei with HS below the shock in
Figs.~\ref{fig:fullstate-bounce} and \ref{fig:fullstate-pb}, but for HS it
persists for a much longer time. Furthermore, in STOS these nuclei are
superheavies with mass numbers up to 500. In general we find that neutrinos are trapped
under the conditions where superheavies appear, and thus they do not have an
effect on the neutrino transport. Therefore we do not expect that the limitation
of HS to nuclei with mass numbers up to $\sim 300$ is crucial for the supernova
dynamics.

We turn back to the evolution of the nuclear distribution of the mass shell
shown in Fig.~\ref{fig:iso}. At bounce, the shock forms very close to
0.6~M$_{\odot}$. Almost immediately after the stage depicted in
Fig.~\ref{fig:iso}(c), the shock runs through the selected mass shell and most
of the heavy nuclei get dissociated (Fig.~\ref{fig:iso}(d)). A distribution of
light nuclei forms, with mass fractions which decrease exponentially with
increasing $A$ (Fig.~\ref{fig:xa0.6a}). This form of the distributions at 1 and
50~ms post bounce indicates that the binding energies are not very important any
more, due to the high temperatures. When the shock propagates outward and
dissociates the infalling matter, the distributions change from the very broad
shape with peaks due to binding energy and shell effects to a simple
exponential. The change of the shape of the distributions can also not be
captured by the average of the heavy nuclei. For the higher mass shells shown in
Figs.~\ref{fig:xa0.6b} and \ref{fig:xa0.6c}, similar features are observed but
at later times. Furthermore, the mass numbers reached are significantly smaller
for the outer mass shells, because matter is much less compressed. Animated
results of the nuclear distributions of different mass shells can be found at
the aforementioned HS EOS Web site.

Also in the post-bounce evolution light nuclei give an important contribution to
the composition, as can be seen from Figs.~\ref{fig:comp-s15p-a} and
\ref{fig:comp-s15p-b}. Again, the standard particles which are usually
considered in a core-collapse supernova are shown in black, and the additional
new light nuclei in blue. One can identify three different regions: the core of
the newly born proto-neutron star at the center up to $R \sim 10$~km, the
envelope of the proto-neutron star up to the standing accretion shock at $R \sim
200$~km, and the infalling matter in the outer layers. The core is composed of
mainly free nucleons and some light nuclei which disappear at densities larger
than saturation density. The accreted matter is a mixture of mainly heavy
nuclei, alpha particles and free protons at low abundances. However, behind the
standing accretion shock the sudden increase of temperature dissociates heavy
nuclei into smaller fragments, a mixture of nucleons and light nuclei appears.
Note that $^3$H is the most abundant light nucleus in the proto-neutron
star, whereas $^2$H dominates the shock heated low-density matter which is
more symmetric. The $^2$H fraction is usually at least one order of magnitude
larger than the alpha-particle fraction.
\subsection{Impact on Neutrino Heating and Cooling}
The presence of light nuclei can modify neutrino heating and cooling. In the present article, we do not include any inelastic weak processes on light nuclei, but the detailed knowledge of their abundances gives already first insight into the possible impact they might have on neutrino heating and cooling. 
The heating region is located between the neutrinospheres and the standing bounce
shock, whereas cooling occurs around the neutrinospheres. To identify the
abundances of light nuclei in the two regions, in Figs.~\ref{fig:comp-s15p-a}
and \ref{fig:comp-s15p-b} we also show the neutrinospheres (see the vertical
lines, $\nu_e$: solid lines, $\bar\nu_e$: dashed lines). During the earlier
post-bounce phase (see bottom panel of Fig.~\ref{fig:composition-light}), $^2$H
and $^3$H are as abundant as protons right behind the still expanding bounce
shock. There, the inclusion of
inelastic weak processes with light nuclei, i.e., inelastic scattering and
most dominantly charged-current reactions, may modify the neutrino heating. 
However, already at 50~ms (Fig.~\ref{fig:comp-s15p-a}) the
abundances directly below the shock have decreased and the neutrino flux is
already quite diluted geometrically at this distance close to the shock. The
most efficient heating occurs at smaller radii, where there are much less light
nuclei than free protons. 

\begin{figure}
\includegraphics[width=\columnwidth, clip=true]{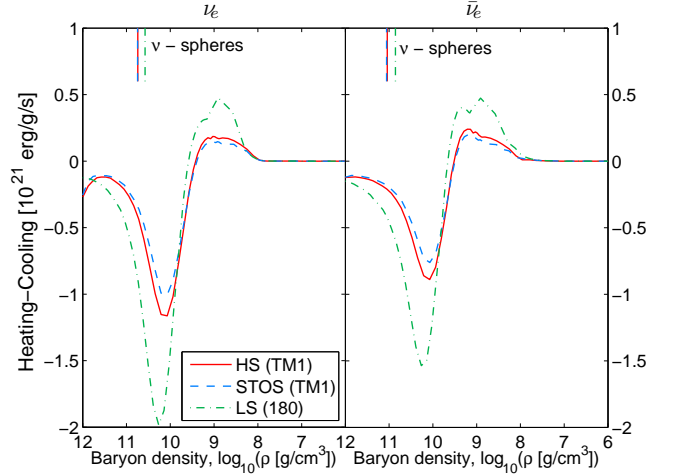}
\caption{Heating--cooling rates, i.e., the net energy deposition by neutrinos for
the 15~M$_\odot$ progenitor models at 200~ms post bounce,
comparing the different EOS under investigation. The vertical lines show the
neutrinospheres.}
\label{fig:heatplot}
\end{figure}

\begin{figure*}[ht]
\subfigure[\ Luminosities]{
\includegraphics[width=\columnwidth, clip=true]{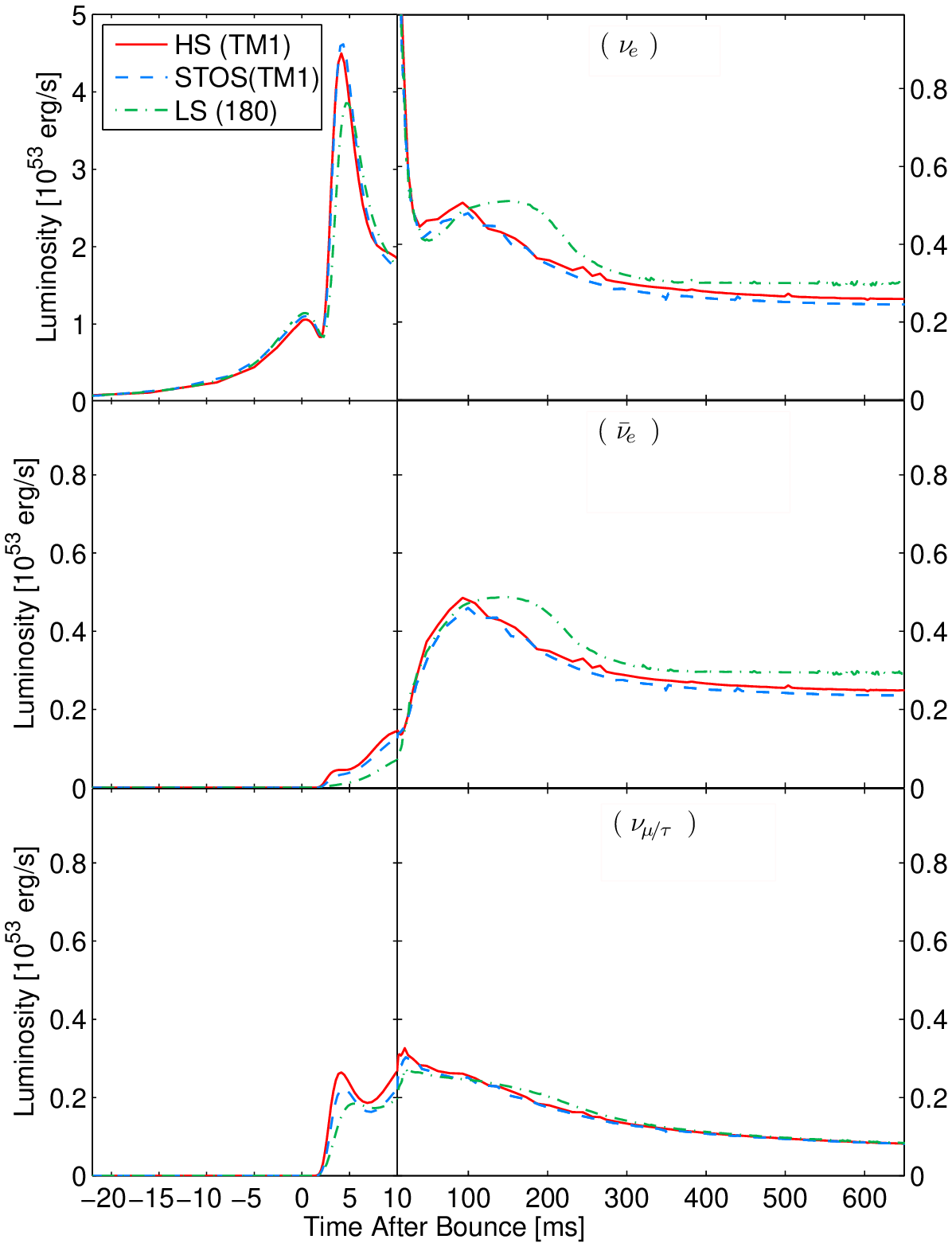}
\label{fig:lumin-s15}}
\hfill
\subfigure[\ Mean energies]{
\includegraphics[width=\columnwidth]{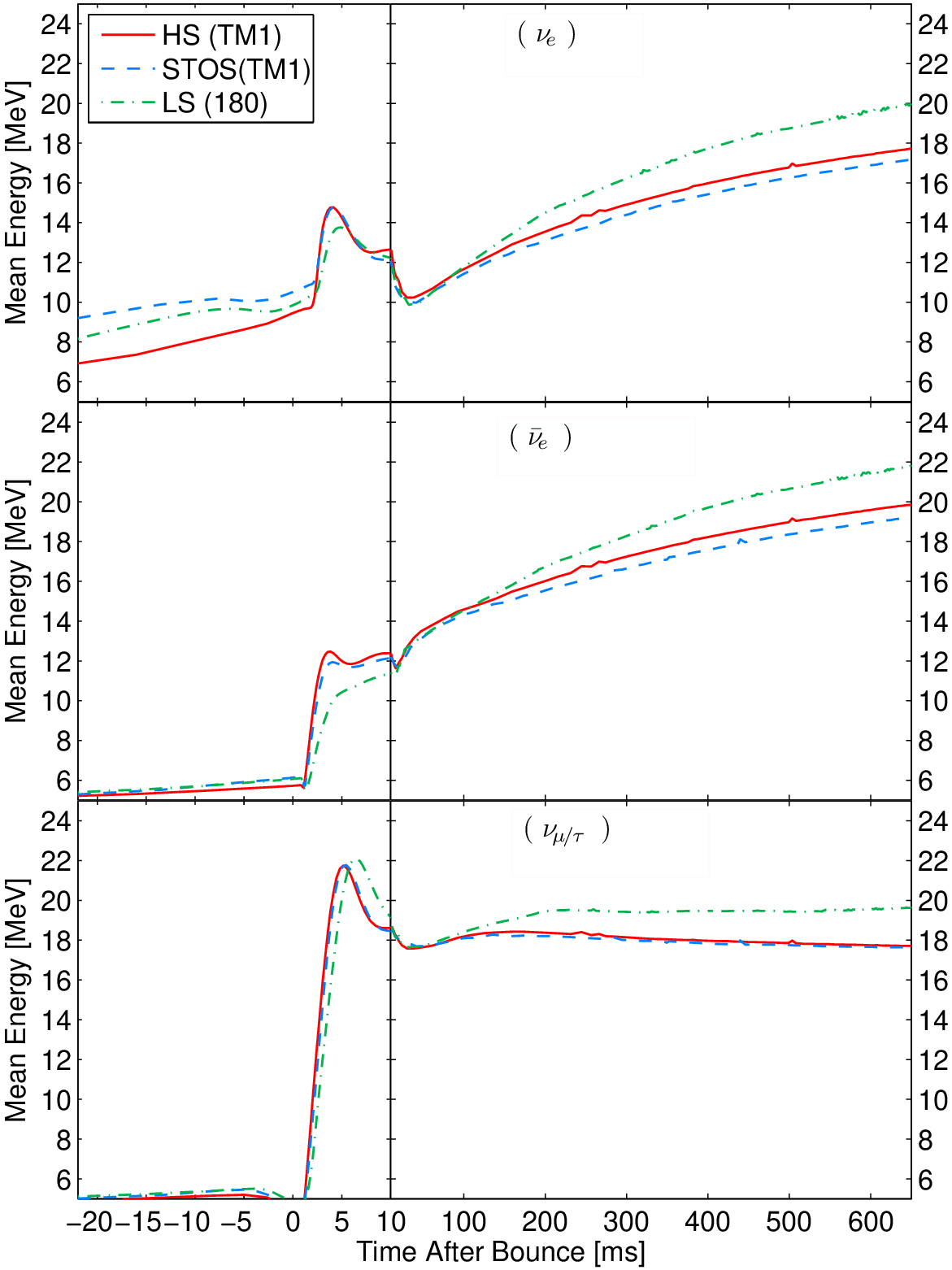}
\label{fig:rms-s15}}
\caption{Evolution of the neutrino luminosities and root-mean-square energies
for the 15~M$_\odot$ progenitor model from \citet{WoosleyWeaver:1995}, comparing
the HS~(TM1) EOS (red solid lines), the STOS~(TM1) EOS (blue dashed lines), and
the LS EOS (dash-dotted green lines). The observables are sampled in a co-moving
reference frame at a distance of 500~km. Note the two different scales in the
subfigure for the electron neutrino luminosities.}
\label{fig:neutrino15}
\end{figure*}

The abundance of light nuclei in the heating region decreases further on the
long term. The slight compression of the surface of the PNS which is seen in
Figs.~\ref{fig:comp-s15p-a} and \ref{fig:comp-s15p-b} decreases the density
close to the shock, leading to a dissolution of light nuclei into nucleons. At
200~ms post bounce, above the neutrinospheres only few light nuclei are found in
comparison to the free nucleons. Hence a very strong impact on neutrino heating
is not expected during the later post-bounce phases. Nevertheless it is
interesting to note that in Fig.~\ref{fig:comp-s15p-b} the largest difference
between the alpha-particle fraction and the fractions of additional light nuclei
is observed in the heating region between the neutrinospheres and the standing
bounce shock. There, the fraction of alpha-particles is up to three orders of
magnitude smaller than the fraction of deuterium. The largest contributions of
light nuclei in the post-bounce evolution are found below the neutrinospheres.
There, e.g., deuterium is still as abundant as protons. It may modify cooling
and could have an effect on the explosion dynamics. For example, less cooling has
a similar potential than more heating. Additionally, an impact on cooling is
interesting because it modulates the neutrino signal. Furthermore, it may be of
particular interest for studies that investigate the long-term evolution of
proto-neutron stars, during which the neutrinospheres move continuously to
higher densities where light nuclei are more abundant. The conclusions about the
possible impact of light nuclei on the supernova dynamics are only qualitatively
and should be investigated in more detail (including inelastic weak processes with light
nuclei) in future studies.
 
Apart from the composition also the EOS affects the heating and cooling via the
compactness of the proto-neutron star. This is shown in Fig.~\ref{fig:heatplot}
where we plot the heating--cooling rates for $\nu_e$ (left panel) and $\bar\nu_e$
(right panel) with respect to the baryon density, at 200~ms post bounce. It
becomes clear that the largest heating below the standing bounce shock (which is
located around $\rho\simeq 10^8$~g/cm$^3$) but also the largest cooling rates at
the neutrinospheres are obtained for the soft LS~(180) EOS. Although STOS~(TM1)
and HS~(TM1) are both based on the same RMF parameterization TM1, the slightly
more compact proto-neutron star of HS leads to larger heating and cooling rates
for HS.
\subsection{Neutrino Signal}
Figures~\ref{fig:comp-s15p-a} and \ref{fig:comp-s15p-b} show that neutrons and
protons are the most abundant particles at the neutrinospheres, but also $^2$H
and $^3$H appear at non-negligible abundances of $10^{-1}$ to $10^{-2}$.
Consequently, the inclusion of the weak processes with light nuclei which
are not incorporated in the present study may modify the neutrino signal.
Here, we will identify the impact of the three EOS on the neutrino
signal. Several aspects related to the comparison of the neutrino signal of STOS
and LS have already been discussed in
\citet{Sumiyoshi:etal:2008} and \citet{Fischer:etal:2009} for various massive iron-core
progenitor models from different groups. Here we will add the differences and
similarities between HS and STOS as well as LS to the discussion.

Figure~\ref{fig:neutrino15} depicts the neutrino signal, i.e., the evolution of
luminosities and mean energies, for the 15~M$_{\odot}$ progenitor comparing the
three reference EOS under investigation. Note the two different scales in the
sub-figure for the electron neutrino luminosity (top panel of
Fig.~\ref{fig:lumin-s15}). Let us first look at the collapse phase before bounce
during which only $\nu_e$ are produced via electron captures at heavy nuclei and
free protons. The earliest time which is shown in Fig.~\ref{fig:neutrino15}
corresponds to a stage between Fig.~\ref{fig:fullstate-collapse1} ($t_{\rm pb} \sim
-40$~ms) and Fig.~\ref{fig:fullstate-collapse2} ($t_{\rm pb} \sim -1$~ms). If we
look at the top panel of Fig.~\ref{fig:rms-s15} we see that the mean energies of
electron neutrinos are about 1.5~MeV smaller comparing HS with LS and about
2~MeV smaller comparing HS with STOS. This is related to the conditions at
decoupling at the neutrinospheres $R_{\nu}$ during the core-collapse phase,
where we find the following ordering: $\rho_{R_{\nu_e}}^{\text{HS}} >
\rho_{R_{\nu_e}}^{\text{STOS}} \geq \rho_{R_{\nu_e}}^{\text{LS}}$,
$T_{R_{\nu_e}}^{\text{HS}} \geq T_{R_{\nu_e}}^{\text{LS}} >
T_{R_{\nu_e}}^{\text{STOS}}$, and $R_{\nu_e}^{\text{HS}} <
R_{\nu_e}^{\text{STOS}} \simeq R_{\nu_e}^{\text{LS}}$. Since low-energy
neutrinos decouple generally at high densities and temperatures with small
neutrinosphere radii, the following hierarchy holds for the mean energies
$\langle E \rangle_{\nu_e}^{\text{HS}} < \langle E
\rangle_{R_{\nu_e}}^{\text{LS}} < \langle E \rangle_{R_{\nu_e}}^{\text{STOS}}$.
The electron neutrino luminosities which are also shown in
Fig.~\ref{fig:lumin-s15} are determined by mass accretion. During the collapse
phase the differences in the mass accretion rates are small for the different
EOS and the ordering of the neutrinospheres leads to the following hierarchy for
the luminosities $L_{\nu_e}^{\text{HS}} < L_{\nu_e}^{\text{STOS}} \leq
L_{\nu_e}^{\text{LS}}$. During the first 10~ms after bounce, when the shock runs
through the neutrinospheres, the $\nu_e$-deleptonization burst is released. We
find that the luminosities and mean energies of LS are lowest with maximum
$L_{\nu_e} \simeq 4\times10^{53}$~erg/s, whereas HS and STOS are very similar
with maximum $L_{\nu_e} \simeq 4.5\times10^{53}$~erg/s.

During the later post-bounce evolution, $\nu_e$ and $\bar\nu_e$ are produced
mainly via electron captures at protons and positron captures at neutrons. Their
luminosities are given by mass accretion at the corresponding neutrinospheres.
The following hierarchy holds for the mass accretion rates at the
neutrinospheres $\dot{m}_{R_{\nu_e/\bar\nu_e}}^{\text{LS}} >
\dot{m}_{R_{\nu_e/\bar\nu_e}}^{\text{HS}} \geq
\dot{m}_{R_{\nu_e/\bar\nu_e}}^{\text{STOS}}$.  Note also the ordering of the
neutrinosphere radii, $R_\nu^\text{LS}<R_\nu^\text{HS}<R_\nu^\text{STOS}$. The
different mass accretion rates are related to the different post-bounce
conditions obtained for the different EOS. LS leads to the most compact
configuration where the standing bounce shock contracts fastest which in turn
leads to the highest mass accretion rate. The opposite holds for STOS, and HS
lies between LS and STOS. Note that the small differences between HS and STOS
are due to the different description of light and heavy nuclei at low and
intermediate densities, which leads to slightly lower $Y_e$ and a slightly more
compact configuration for HS. Hence,
$L_{\nu_e/\bar\nu_e}^{\text{LS}} > L_{\nu_e/\bar\nu_e}^{\text{HS}} \geq
L_{\nu_e/\bar\nu_e}^{\text{STOS}}$ throughout the entire post-bounce phase shown
in Fig.~\ref{fig:lumin-s15} up to 650~ms. On the other hand, $\nu_{\mu/\tau}$
are produced by pair processes and they interact only via neutral--current
reactions. They decouple at generally higher densities. Consequently their
neutrinospheres are deeper inside the proto-neutron star than the electron
flavor neutrinospheres. The $\nu_{\mu/\tau}$ luminosity can be expressed in
terms of diffusion which scales as
$R_{\nu_{\mu/\tau}}^2\,T^4\vert_{R_{\nu_{\mu/\tau}}}$. Here, higher temperatures
at the neutrinospheres are compensated by smaller neutrinosphere radii and hence
the luminosities have similar values for all EOS under investigation during the
post-bounce evolution shown in Fig.~\ref{fig:lumin-s15} up to 650~ms. 

The ordering of the mean energies of the different EOS is the same as for the
luminosities. It reflects the ordering of the temperatures at the
neutrinospheres. However, mean energies of $(\mu/\tau)$-neutrinos for LS are
significantly higher than for HS and STOS. Note that $(\mu/\tau)$-neutrinos are
sensitive to differences at high densities, where the EOS under investigation
lead to different proto-neutron star contraction behaviors. The significantly
faster proto-neutron star contraction for LS, in comparison to HS and STOS,
leads to the higher  $(\mu/\tau)$-neutrinos mean energies shown in
Fig.~\ref{fig:rms-s15}. The contraction behavior for HS and STOS is more similar
at high densities, leading to almost equal mean  $(\mu/\tau)$-neutrino energies.

\begin{figure*}
\subfigure[\ Luminosities]{
\includegraphics[width=\columnwidth, clip=true]{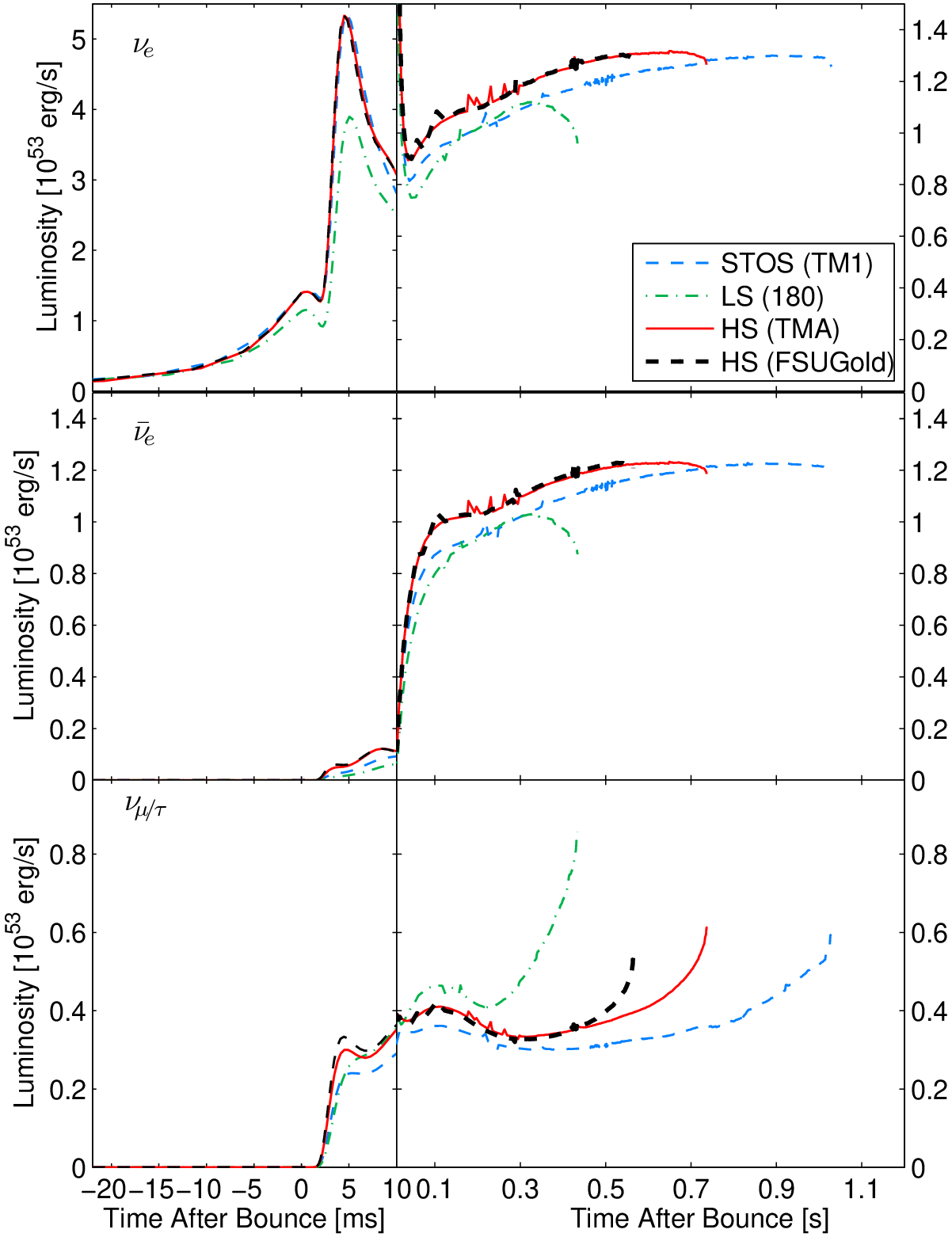}
\label{fig:lumin-s40}}
\hfill
\subfigure[\ Mean energies]{
\includegraphics[width=\columnwidth, clip=true]{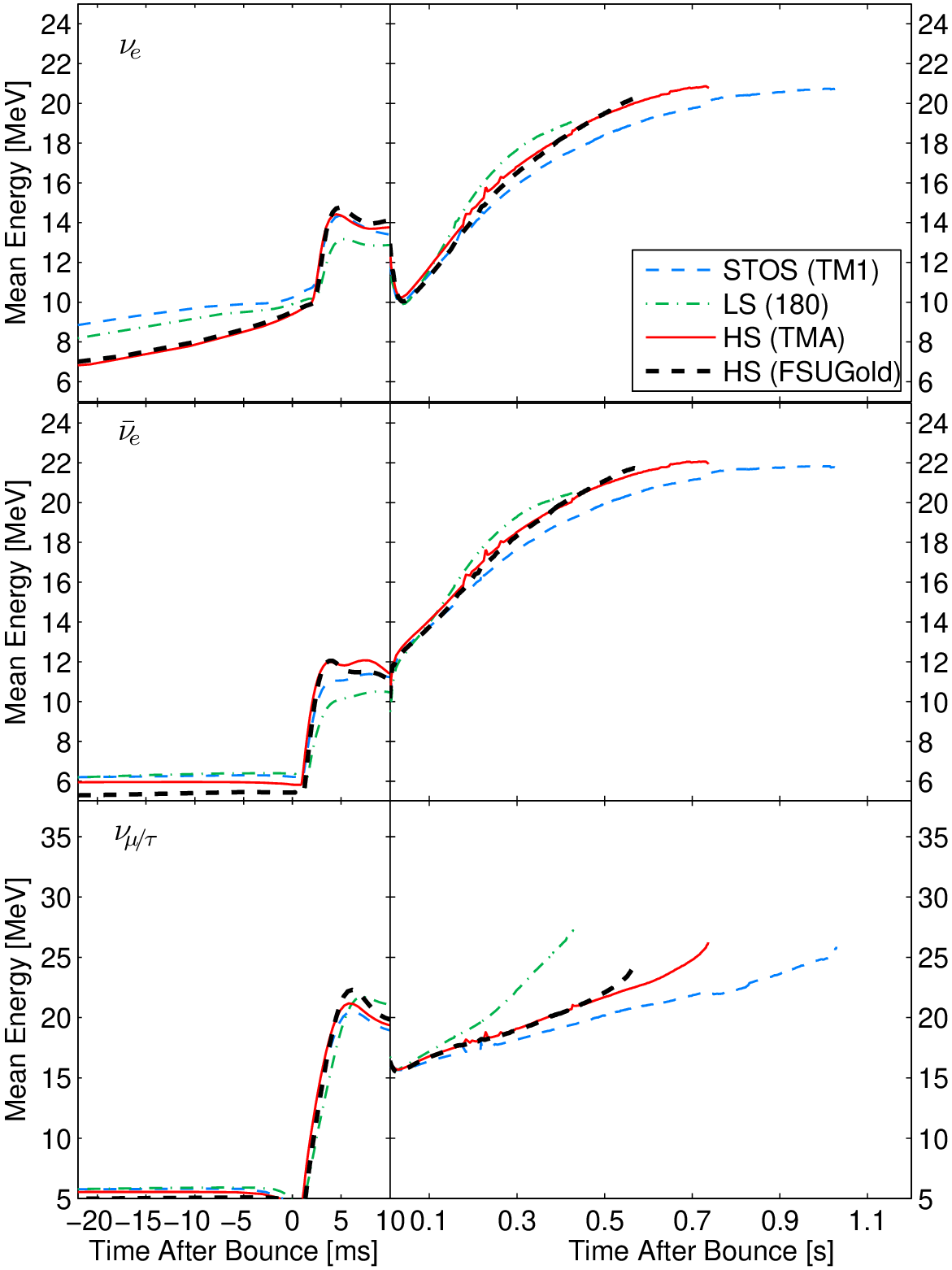}
\label{fig:rms-s40}}
\caption{Evolution of the neutrino luminosities and mean energies for the
40~M$_\odot$ progenitor model from \citet{WoosleyWeaver:1995}, comparing the HS
(TMA) EOS (red solid lines), HS (FSUgold) EOS (black dashed lines), the
STOS~(TM1) EOS (blue dashed lines), and the LS~(180) EOS (dash-dotted green
lines). The observables are sampled in a co-moving reference frame at a distance
of 500~km. Note the two different scales in the subfigure for the electron
neutrino luminosities.}
\label{fig:neutrino40}
\end{figure*}
\section{Results: different nuclear interactions}
\label{sec_40}
The EOS imprint on the neutrino signal for the 15 M$_{\odot}$ progenitor
is rather weak. However, if we go to the regime of failed supernovae of very
massive progenitors, the influence of the EOS becomes more pronounced, as was
also found in previous EOS studies, like, e.g., by
\citet{Sumiyoshi:etal:2007}, \citet{Fischer:etal:2009}, and \citet{oconnor11}. In order to illustrate
and understand the effect of the high-density EOS on the emitted neutrino signal
from core-collapse supernovae, we simulate the collapse of the massive
40~M$_\odot$ progenitor from \citet{WoosleyWeaver:1995} which eventually ends up
in a black hole. For such a massive progenitor, very high densities and
temperatures are reached, where different EOS show significant differences.
We want to emphasize that this is the first time that the
non-relativistic LS EOS are compared with several different RMF EOS, giving a more comprehensive view of EOS effects. On the other hand, we do
not explore the important sensitivity on the progenitor star which was
thoroughly studied by \citet{oconnor11}.

From the relativistic EOS models we apply STOS~(TM1) and the two new EOS tables
HS~(TMA) and HS~(FSUgold). Because HS~(TM1) and STOS~(TM1) are identical at
high densities, we do not include HS~(TM1) in the comparison any more. The two
new EOS tables HS~(TMA) and HS~(FSUgold) have been introduced in
\S \ref{sec_nse} and \S \ref{subsec_char}, and selected EOS properties are
listed in Table~\ref{nse_table_rmf}. We consider two different versions of the
non-relativistic LS EOS in the comparison, LS~(180) and LS~(220), characterized
by different values of the incompressibility (see Table~\ref{nse_table_rmf}).
However, the simulation with LS~(220) is performed with a lower resolution where
we use only four neutrino propagation angles, because for this model we restrict
the discussion on hydrodynamical aspects.

The most important hydrodynamical quantities at the core bounce of the five
different EOS are given in Table~\ref{tab:bounce40}. If one compares
these values with the EOS properties of Table~\ref{nse_table_rmf}, one sees
that LS~(180) with the lowest incompressibility has the highest central density
at bounce, followed by LS~(220) and then FSUgold which has an incompressibility
of 230~MeV. For TM1 and TMA the correlation between the density at bounce and
the incompressibility is inverted. TM1 has the lowest density, even though the
incompressibility of TMA is the largest. This has two reasons: here, we have
asymmetric nuclear matter and the large symmetry energy of TM1 stiffens the
EOS. Furthermore, higher order terms for the expansion around saturation density
for symmetric matter (e.g., the skewness) are larger in TM1, so that TM1 behaves
stiffer, as also noted before in \S~\ref{subsec_char}. This example
illustrates that it is generally not possible to explain the global properties
of an EOS with single saturation parameters, especially when going to high
densities and/or asymmetry. 

\begin{table}
\caption{Selected Central Hydrodynamic Quantities at Bounce for the 40~M$_\odot$
Models}
\centering
\begin{tabular}{l c c c c}
\hline
\hline
EOS & $\rho$ & $s$ & $Y_e$ \\
& $[10^{14}$ g/cm$^3 ]$ & [k$_B$] &  \\
\hline
LS~(180) & 3.855 & 1.487 & 0.2844  \\
LS~(220) & 3.674 & 1.475 & 0.2826  \\
HS~(FSUgold)  & 3.151 & 1.529 & 0.2815  \\
HS~(TMA)  & 2.878 & 1.545 & 0.2811  \\
STOS~(TM1) & 2.690 & 1.505 & 0.2915  \\
\hline
\hline
\end{tabular}
\label{tab:bounce40}
\end{table}

The most interesting differences occur in the late post-bounce phase, shortly
before collapse to a black hole. Here, one could expect that the maximum
mass of a cold neutron star is correlated with the time until black hole
formation and the corresponding neutrino signal, as was also found in previous
studies of \citet{Sumiyoshi:etal:2007}, \citet{Fischer:etal:2009}, and \citet{oconnor11}. If this
expectation was true, we should find the following sequence for the time until
black hole formation: HS~(FSUgold), LS~(180), HS~(TMA), LS~(220), and STOS~(TM1), in increasing order. Next we investigate the neutrino signal and try
to identify such correlations.

The neutrino signals for the 40~M$_\odot$ model using different EOS are shown in
Fig.~\ref{fig:neutrino40}, similar to Fig.~\ref{fig:neutrino15}.
LS~(220) is not included in this figure, due to the reduced neutrino
angular resolution in its simulation. We remark that the mean energies shown in
Fig.~\ref{fig:rms-s40} are always continuous. However, in the first milliseconds
after bounce, they can change so quickly that this cannot be resolved in the
scaling of the figure. The differences obtained during the core-collapse phase
before bounce for the 40~M$_\odot$ models are equivalent to what we have already
discussed in the previous section at the example of the 15~M$_\odot$ models. For
the two HS models the neutrino light curves and evolution of the mean energies
lie almost on top of each other, independent from the parameterization, until a
certain level of compactness is reached. This also holds for the HS (TM1) EOS
which is not shown here. It is very interesting that all HS EOS give a very
similar neutrino signal even during the early post-bounce phase. The neutrino
signal is generated at densities below saturation density, and even though the
nascent proto-neutron star may be different for the different EOS, the imprint
of the high-density phase is rather weak. Obviously, at low densities the
interactions of unbound nucleons are not so important, whereas the
description of nuclei and the degrees of freedom in the nuclear composition are
crucial, which explains why the neutrino signal evolves so similar for the
different HS models. 

\begin{figure}
\centering
\includegraphics[width=0.9\columnwidth, clip=true]{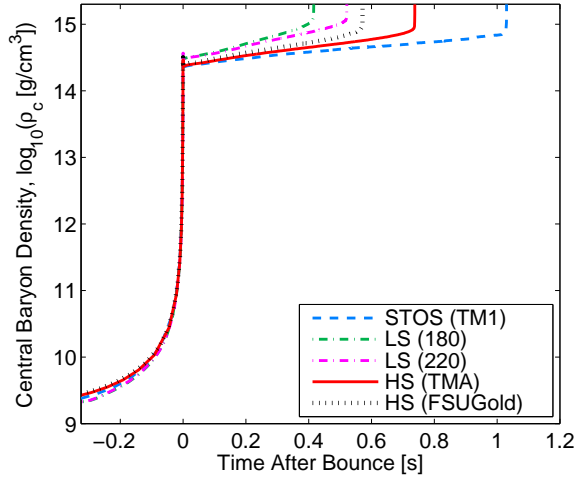}
\caption{Evolution of the central density for the collapse of the 40~M$_\odot$
progenitor model from \citet{WoosleyWeaver:1995}, comparing the different EOS
under
investigation.}
\label{fig:dens_s40}
\end{figure}

However, the moment of black hole formation is set by the nuclear interactions
in the high-density region of the EOS. Figure \ref{fig:dens_s40} depicts the
evolution of the central densities for the five different EOS. After bounce,
the curves first show a linear behavior, until the newly born proto-neutron star
becomes gravitationally unstable and collapses to a black hole. LS~(180) shows a very fast
compression in the center, followed by LS~(220), then FSUgold, then TMA and
finally TM1 with the slowest compression behavior. By approaching
the moment of black hole formation a significant rise of the $\nu_{\mu/\tau}$
luminosities is observed. The different EOS have different timescales until
black hole formation, and thus pronounced differences occur for the
$\nu_{\mu/\tau}$ luminosities and mean energies, making $\mu / \tau$-neutrinos
the best messengers of the EOS. Contrary, there is only little separation for
the evolution of $\nu_e$ and $\bar \nu_e$ mean-energies for the four EOS, and
also the luminosities remain more similar. Further details about the rise of the
$\nu_{\mu/\tau}$ luminosities and its connection to the contraction of the
proto-neutron star can be found in \citet{Liebendoerfer:etal:2004} and
\citet{Fischer:etal:2009}. 

\begin{figure}[t]
\centering
\includegraphics[width=0.8\columnwidth, clip=true]{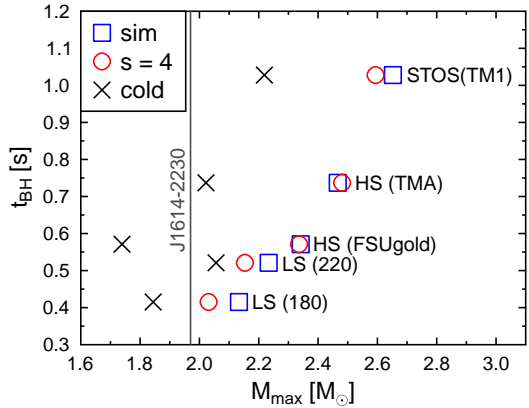}
\caption{Time for the onset of collapse to a black hole for the
40~M$_\odot$ progenitor of \citet{WoosleyWeaver:1995} with respect to different
configurations of maximum gravitational masses for the investigated EOS.
``cold'': static cold neutron stars at $T=0.1$~MeV and beta-equilibrium without
neutrinos. ``$s=4$'': static proto-neutron stars at a constant entropy per
baryon of $s=4$~k$_B$ and beta-equilibrium without neutrinos. ``sim'': the
maximum gravitational masses found in the simulations. The vertical line shows
the mass measurement of pulsar PSR J1614-2230 by \citet{Demorest:etal:2010}.}
\label{fig:tm}
\end{figure}

Table~\ref{tab:bhcoll} lists selected hydrodynamic quantities for
the last stable configuration before the beginning of black hole formation. Here
this is defined as the moment when infall velocities on the order of 1000~km/s
are obtained in the PNS interior. We study the possible correlation between the
maximum mass of a cold neutron star and short/long accretion time until black
hole formation by comparing $t_{\rm pb}$ and $M_G$ from Table~\ref{tab:bhcoll} with
the maximum masses of cold neutron stars given in \S~\ref{subsec_char}. The
results are shown in Fig.~\ref{fig:tm}, where the black crosses depict the time
until beginning of black hole formation as a function of the maximum
gravitational masses of cold neutron stars. The maximum gravitational masses
from Table~\ref{tab:bhcoll} which were found in the simulations are shown by the
blue squares. Compared to the cold configurations they are significantly
increased for all EOS, but not equally strong. Surprisingly, although LS~(180)
has a larger maximum mass for a cold neutron star than FSUgold, and the maximum
mass of LS~(220) is larger than for FSUgold and TMA, the two LS EOS lead to the
earliest beginning of black hole formation. In the supernova environment they
behave less stable than the RMF EOS and even FSUgold gives a later black hole
formation then both of them. FSUgold is followed by TMA and then comes TM1 with
the latest collapse. Consequently, the expected ordering with the maximum mass of cold neutron stars established in
the literature so far is not valid any more for our extended set of EOS
including the new TMA and FSUgold EOS. The hierarchy between the maximum mass of
cold neutron stars and short/long accretion times before black hole formation
holds only separately within the class of RMF models or within the class of the
non-relativistic LS EOS.

We want to examine this result further. In Table~\ref{tab:bhcoll},
the time until black hole formation and the enclosed baryon mass have the same
ordering, which shows that the accretion rate is affected only little by the
EOS. On the other hand the central densities at the onset of collapse are in
most cases lower for the EOS which have a large mass at a later collapse. It illustrates the
stiffness of the EOS for the conditions encountered here. Interestingly, for the
central temperatures at the onset of collapse to a black hole, which are
listed in Table \ref{tab:bhcoll}, we find roughly 50~MeV for all three RMF EOS,
but only 30~MeV for the two non-relativistic LS EOS.

\begin{figure}
\centering
\includegraphics[width=\columnwidth, clip=true]{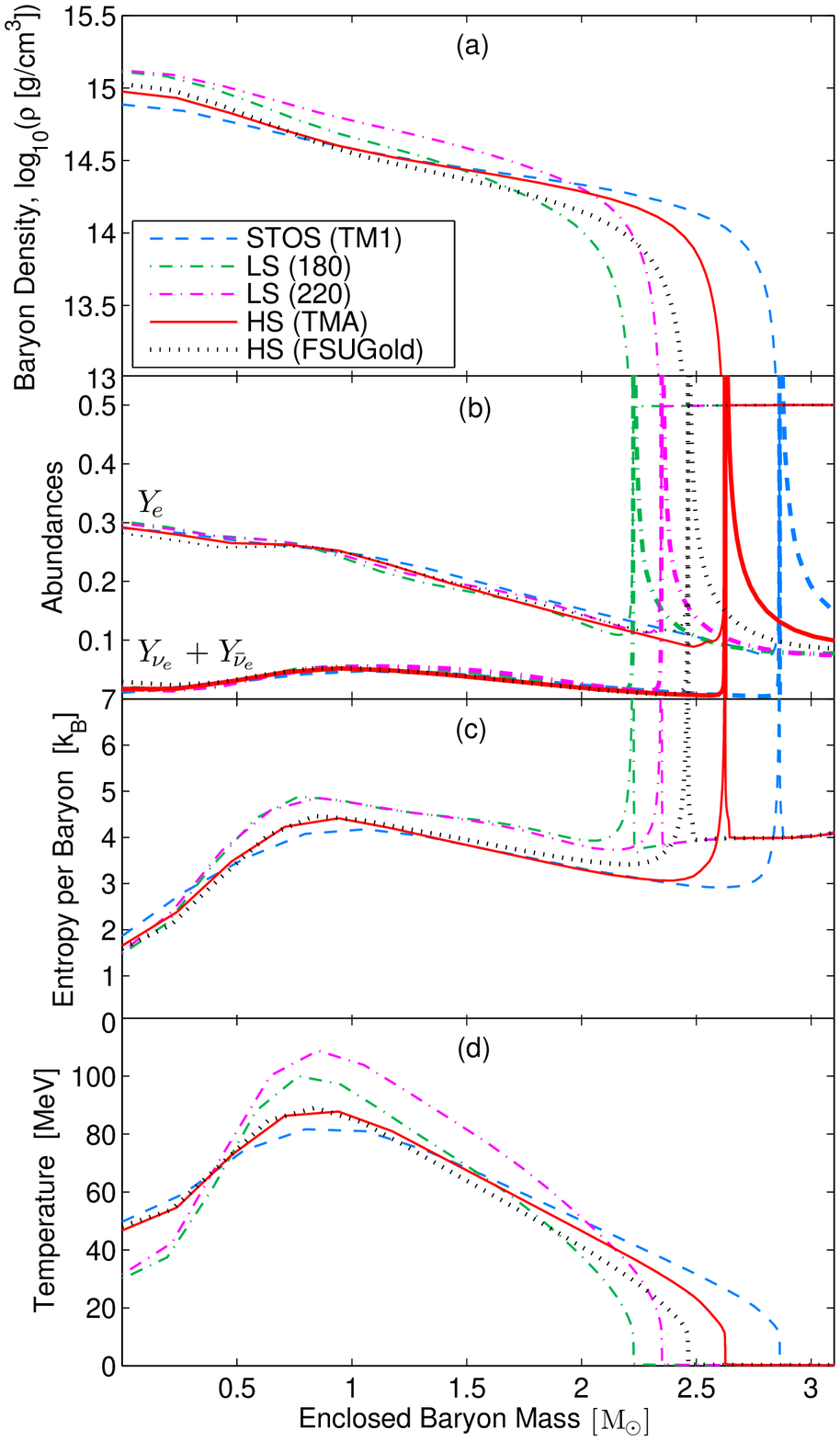}
\caption{Selected quantities at the onset of collapse to a black hole of the
40~M$_\odot$ progenitor model from \citet{WoosleyWeaver:1995}, comparing the different EOS
under investigation.}
\label{fig:hydro-bh}
\end{figure}

The hydrodynamic state corresponding to the last stable
configuration is further illustrated in Fig.~\ref{fig:hydro-bh}. The similar
behavior of the central temperatures of the RMF EOS and the different behavior
of LS can be seen clearly. The peak temperatures around $M=0.8$~M$_{\odot}$ are
different for all five models. Interestingly, they show the same ordering like
the central densities at core bounce in Table~\ref{tab:bounce40}, namely, 
LS~(180), LS~(220), HS~(FSUgold), HS~(TMA), STOS~(TM1) in decreasing order. This
is also the sequence for the time until black hole formation. Note that the same
ordering is also seen in the luminosities and mean-energies of
$\mu/\tau$-neutrinos, which are most sensitive to temperature effects. The
entropy profiles look more similar than the temperature profiles, whereas the
two LS EOS lead to slightly higher entropies for $M_B>0.5$~M$_{\odot}$. In
Fig.~\ref{fig:hydro-bh} one can also identify the different central densities
and different masses of the proto-neutron stars mentioned before. The electron
fraction profiles show only small differences for the investigated EOS. The
neutrino abundances are rather small and are also very similar for the different
EOS.

\begin{table}
\caption{Selected Quantities at the Onset of Collapse to a Black Hole}
\centering
\begin{tabular}{l c c c c c}
\hline
\hline
EOS & $t_{\rm pb}$ \footnote{Time post bounce.} & $\rho$ \footnote{Baryon density in
the center.} & $T$ \footnote{Temperature in the center.} & $M_B$ \footnote{Baryon
mass enclosed inside the shock.} & $M_G$ \footnote{Gravitational mass enclosed
inside the shock.}
\\
& $[s]$ & $[10^{15}$ g/cm$^3 ]$ & $[MeV]$ & $[$M$_\odot]$ & $[$M$_\odot]$
\\
\hline
LS (180)       & 0.415 & 1.292 & 29.978 & 2.227 & 2.133 \\
LS (220)       & 0.521 & 1.324 & 31.446 & 2.350 & 2.233 \\
HS (FSUgold)   & 0.571 & 1.058 & 48.104 & 2.465 & 2.341 \\
HS (TMA)       & 0.737 & 0.943 & 46.708 & 2.626 & 2.466 \\
STOS (TM1)     & 1.028 & 0.769 & 49.705 & 2.864 & 2.652 \\
\hline
\hline
\end{tabular}
\flushleft
{\bf Notes:}
\label{tab:bhcoll}
\end{table}

We come to the unsatisfying conclusion that neither the
incompressibility (comparing HS~(TMA) and STOS~(TM1)) nor the maximum masses of
cold neutron stars can directly be related to the time until black hole
formation. Single saturation properties are not sufficient to describe the
behavior of the EOS in the simulation, where the different EOS evolve to
different thermodynamic states. The maximum mass of cold neutron stars is not
very meaningful, because the found states are very different from the cold
configurations and because the strength of the temperature effects is model
dependent as we will discuss in the next paragraph.

The profiles at the beginning of black hole formation from the simulations shown in Fig.~\ref{fig:hydro-bh} can roughly be approximated by a constant entropy per baryon of $s = 4$~k$_B$ and electrons in beta-equilibrium, where the contribution of neutrinos is negligible. We can compare the results from the simulations with hydrostatic configurations of the Tolman--Oppenheimer--Volkoff equations by using these approximations.
The corresponding maximum masses are plotted in Fig.~\ref{fig:tm} versus the time until black hole formation, $t_{\textrm{BH}}$, from Table~\ref{tab:bhcoll}, by red circles. If one compares with the gravitational masses from the simulations which are
depicted by blue squares, one sees that there is a rough agreement between
``$s=4$'' and ``sim''. Furthermore, the time until black hole formation is
monotonically increasing with $M_\textrm{max}(s=4)$, similar as for
$M_\textrm{max}(\textrm{sim})$. This shows that the above approximations ($s = 4$~k$_B$, beta-equilibrium without neutrinos) work sufficiently well to describe the states at the beginning of black hole formation from the simulations. In conclusion, the significant increase of the maximum
mass in the simulation and the weak correlation between $M_\textrm{max}(\textrm{cold})$
and $t_{\textrm{BH}}$ is explained mainly as a temperature effect. 

Very often for hydrostatic configurations of proto-neutron stars a constant entropy per baryon of $s =2$~k$_B$ and electron lepton fraction $Y_L = 0.4$ is assumed. These conditions are clearly different to the profiles shown in Fig.~\ref{fig:hydro-bh} and are more appropriate for proto-neutron stars of less massive progenitors. We remark that such hydro-static configurations with $s =2$~k$_B$ and $Y_L = 0.4$ are indeed not suitable to explain the observed behavior. Their maximum masses are 0.3--0.5 M$_\odot$ lower than observed in the simulations, and do not show a monotonic correlation with $t_{\textrm{BH}}$. A large entropy of $s = 4$~k$_B$ is the important point for the explanation, whereas the neutrino fraction is less significant. 

It is interesting that
the non-relativistic LS EOS show less stiffening at finite entropy than the
relativistic EOS. The different temperature profiles discussed before also
indicate a different temperature dependency of the LS EOS.
It is difficult to pin down whether this is an artifact of the
non-relativistic dispersion relation or just a result of the chosen
parameterization of the nuclear interactions. For example, the
temperature--entropy relation depends on the effective nucleon mass $m^*$ which appears in the dispersion relation. In LS
the effective mass is set equal to the vacuum mass, $m^*/m=1$, whereas $m^*/m$
goes down to 0.2 in the core of the proto-neutron stars for the RMF models. The
neglect of the effective mass reduction at finite density in LS leads to
a suppression of temperature effects. In addition to the effective mass, also
the electron fraction affects the temperature--entropy relation, because the
electron contribution depends differently on temperature than the nucleon
kinetic contribution. The electron fraction on the other hand is set by the
symmetry energy. These effects, which are independent from the used
energy--momentum relation, can be identified by comparing different RMF models or
the two LS EOS. For example, LS~(220) shows less dependence on entropy than
LS~(180). The same result was found by \cite{oconnor11}. It could be due to the
dominance of the LS~(220) EOS by interactions compared to the kinetic
contribution. For the RMF models we see that FSUgold has the
largest increase of the maximum mass at constant entropy. This can be related to
its softer density dependence of the symmetry energy at high densities.

Independent of the detailed origin, with these results we
have found a quantity which is directly correlated with the time until black
hole formation: the maximum mass of proto-neutron stars, approximated by the
``$s=4$'' configuration. The time until black hole formation is monotonically
increasing with the maximum masses of ``$s=4$'' proto-neutron stars with an approximately linear dependency. If for a known progenitor one could
detect and follow the neutrino signal of the core-collapse supernova until black
hole formation where the neutrino signal ceases, the correlation of
Fig.~\ref{fig:tm} would give information about the maximum masses of such ``$s=4$''
configurations. This would represent a significant constraint for the nuclear EOS at finite entropy, complementary to observations of cold neutron stars.

The mass measurement of PSR J1614-2230 by \citet{Demorest:etal:2010} was a
breakthrough for the physics of neutron stars. If a more massive pulsar was
detected in the future, the constraints would become even more severe. However,
pulsar mass measurements give always only a lower limit for the actual maximum
mass. Contrary, with the black hole formation one has direct access to the
(proto-neutron star) maximum mass. Furthermore, if we would know the real
maximum mass of cold neutron stars, then one could use the neutrino signal to
constrain the temperature dependence of the EOS. On the other hand, the
correlation of Fig.~\ref{fig:tm} has also a lot of predictive power if it is
combined with the mass constraint of pulsar PSR J1614-2230. All EOS with black
crosses to the left of this line are not compatible with the observational data
of PSR J1614-2230, because their maximum masses of cold neutron stars are too
low. Thus we expect that the neutrino signal of such a black hole formation
event from a massive 40~M$_\odot$ progenitor will not be much shorter than
0.5~s. On the other hand, if the signal would end at $t_{\rm pb}\ll 0.5$~s this
would indicate the missing of important physics in the current models.

Finally, we have to make a remark of caution for the results of this section:
for such high temperatures as encountered here, extensions of the EOS may
become important, see, e.g., \citet{oertel10}. Temperatures of 100~MeV are on the
order of the pion mass, so a pion gas would form, and also muons and baryon
resonances could give an important contribution to the EOS. In addition one has
to be cautious using a non-relativistic EOS at such extreme conditions. It is
not warranted that the EOS remains causal, i.e., that the speed of sound stays
below the speed of light. We find that LS becomes acausal during the collapse to
a black hole. However, we do not expect that our results will change
qualitatively by these aspects, but still they should be studied more elaborately with extended EOS in the future.

\section{Summary and conclusions}
\label{sec_sum}
We apply the new RMF EOS of the statistical model of
\citet{HempelSchaffnerBielich:2010} (HS) to core-collapse supernova simulations
of 15 and 40~M$_\odot$ progenitors. Our core-collapse model is based on general
relativistic radiation hydrodynamics that employs three-flavor Boltzmann
neutrino transport. The HS EOS include up to 8000 nuclei, based on
experimentally known masses of \citet{Audi:etal:2003} and different theoretical
nuclear structure calculations. Three different parameterizations of RMF
interactions are applied, TM1, TMA, and FSUgold which govern the high-density
behavior of the EOS. Regarding the nuclear matter properties and
mass--radius relation of cold neutron stars, none of the HS EOS used here was
able to fulfill all of the quoted experimental and observational constraints.
The same holds for the commonly used EOS from H.~\citet{Shen:etal:1998} (STOS) and
the EOS of \citet{LattimerSwesty:1991} (LS) with an incompressibility of 180~MeV
(LS~(180)). Furthermore, we also include the LS EOS with an incompressibility of 220 MeV (LS~(220)), which is in reasonable agreement with the considered constraints.
The HS EOS tables are available for use and can be accessed from the home page
given in footnote~\ref{eospage}. Also routines are available which allow to
calculate the abundances of all considered nuclei. Additional EOS tables for the
NL3 \citep{Lala97} and DD2 \citep{Typel:etal:2010} RMF parameterizations can also be
downloaded, which were not included in the present study. We remark that DD2 is
in good agreement with the quoted experimental and observational constraints. In
the future, the EOS tables will also be provided on the CompOSE(db) online
platform of the CompStar research networking program.

We identify the most
important differences between the HS EOS with the TM1 parameterization, the STOS
EOS which is also based on TM1, and the LS~(180) EOS by simulating the evolution of the 15~M$_\odot$ progenitor. For the first time, two EOS are available which have identical nucleon interactions but a different description of the low-density phase where nuclei exist.
In general
one of the big advantages of the HS EOS, in comparison to LS and STOS, is the
detailed information of the nuclear composition, the precise description of
nuclei including shell effects, and the consideration of additional light nuclei.
Regarding the first two aspects, the low-temperature HS EOS leads to a smooth
transition from NSE to non-NSE at temperatures
of about 0.5~MeV. There, HS matches the ideal gas of $^{56}$Fe or $^{56}$Ni
(depending on the proton-to-baryon ratio) by construction, while LS and STOS
produce artificial jumps, e.g., in internal energy. Also the
agreement with the composition and the thermodynamic state of the progenitor was
most satisfactory with HS. In the same sense, the
HS EOS is well suited to be connected to explosive nucleosynthesis calculations, as it naturally gives a realistic seed distribution. Obviously,
this is true for all NSE models which employ experimental nuclear binding energies, but the HS EOS tables also cover all densities and
temperatures relevant for supernova simulations, which is not typical for most
of the NSE models.

During the collapse we observe that HS (TM1) gives significantly smaller nuclei
than STOS~(TM1), leading to a stronger deleptonization and a lower electron
fraction in the core. Contrary, the lower free proton fraction in the outer
layers using HS caused less electron captures there. This illustrates
impressively the sensitivity of the core-collapse simulation to slight changes
of the nuclear composition. The electron captures in HS are less energetic,
partly due to the use of real nucleon masses in the EOS, which therefore
contain the proton-to-neutron rest-mass difference. During the collapse HS
leads therefore to less entropy generation in the center compared with STOS. 

The collapse phase is dominated by heavy nuclei. We investigated the nuclear
distributions and identified a visible effect of neutron magic shells in the
nuclear composition, leading, e.g., to several peaks in the mass distributions,
which cannot be captured by the SNA. Interestingly, closed
neutron shells also have an influence on the electron fraction profiles. In the
innermost zones the nuclear distributions become very broad during collapse,
due to the large compression heating. It would be a very appealing extension of
the HS EOS model to calculate weak reaction rates with the nuclear distributions
of the EOS. Here, we only considered the average heavy nucleus in combination
with the reaction rates of \citet{Bruenn:1985}. 

The appearance of additional light nuclei, which so far were not
implemented in EOS tables for core-collapse supernovae, was investigated in detail. The formation of weakly bound light nuclei is driven by entropy.
They are formed in the supernova by an interplay of density and temperature, where the temperature has to be high enough to suppress the role of differences in binding energies. 
We find that during the early collapse phase, the commonly used approximation to
consider only alpha particles of all possible light nuclei is acceptable.
However, already a few milliseconds before bounce temperatures are sufficiently high so that additional light nuclei can be
even more abundant than alpha particles. When the shock runs through the
infalling matter the well-known dissociation of heavy nuclei occurs. In the
present article we further show that a significant fraction of light nuclei,
e.g., $^2$H, $^3$H, $^3$He, and Li, forms in the hot dissociated matter between the
shock and the core of the proto-neutron star. These light nuclei remain there
also in the post-bounce phase. For the inner layers of the proto-neutron star
envelope, where matter is very asymmetric, tritium gives the largest
contribution. In the outer layers below the shock, deuterium plays the most
important role. We remark that these results for light nuclei obtained with the HS EOS 
are in qualitative agreement with expectations from the quantum many-body approaches of \citet{roepke11} and \citet{Typel:etal:2010} (see \citet{hempel11b}) and the previous study of \citet{SumiyoshiRoepke:2008}.

In the present
article, we simplified the neutrino reactions with light nuclei by treating all
of them as alpha particles, i.e., we only considered elastic neutral--current reactions. However, light nuclei have low binding energies,
except alpha particles. Because of the lower binding energies we expect that
weak reactions with the other light nuclei will be different compared to the
weak reactions with alpha particles. Some aspects may resemble the reactions
with free nucleons. For example, also electron captures or break-up reactions of
the deuteron are likely to occur. Until 50~ms post bounce the abundances of deuterons and tritons
close to the shock are comparable to the ones of free protons. Thus. there may be
an influence of light nuclei on neutrino heating during the early post-bounce
phase. However, the fractions of light nuclei in the heating region decrease in
the subsequent evolution. The largest fractions of light nuclei are then found
below the neutrinospheres. Thus for the later post-bounce evolution cooling may
be affected rather than heating. More detailed studies are required to answer to
which extent light nuclei can contribute to the neutrino heating and cooling and
possibly on the explosion dynamics. We see the consistent implementation of weak
processes with light nuclei in the neutrino transport of core-collapse supernova
simulations as an interesting task for future studies. 

Regarding the hydrodynamic behavior of the new HS~(TM1) EOS, we observed that it
leads to a more compact proto-neutron star with slightly higher temperatures
than with STOS~(TM1). This is a result of the different description of the
non-uniform nuclear matter phase, giving the appearance of light nuclei and
smaller heavy nuclei, which in turn leads to lower central electron fractions in the simulations. During the post-bounce evolution, the differences
between HS~(TM1), LS~(180), and STOS~(TM1) reduce. However, the much softer
LS~(180) EOS has significantly higher central temperatures and densities, and
hence the standing accretion shock contracts faster. We arrive at the general
conclusion that the model for low-density nuclear matter, where nuclei exist,
and the implemented nuclear degrees of freedom are as important as the
description of the nuclear interactions around and above saturation density. 

We also compare the evolution of neutrino luminosities and mean energies for the different EOS, for the 15 M$_\odot$ model. The soft
LS~(180) EOS leads to high neutrino fluxes and mean energies. Although
STOS~(TM1) and HS (TM1) are based on the same RMF parameterization, the larger
deleptonization for HS results in larger neutrino fluxes and mean energies than
STOS, emphasizing our statement above. The neutrino signatures are related to
the evolution of the neutrinospheres which are located at the proto-neutron star surface. For LS~(180) the proto-neutron star contracts
fastest with highest temperatures and $Y_e$ obtained, while STOS and HS have
significantly longer contraction times especially after 300~ms post bounce.
However, the differences obtained in the neutrino luminosities and mean energies
are rather small. Therefore for such intermediate-mass progenitors, very precise
neutrino observations of a possible future galactic supernova were required to
constrain the nuclear EOS further. Note, this statement remains only valid
if EOS effects are not affected significantly by the explosion, which is expected
to occur for such a progenitor. 

Pronounced EOS effects are found for the neutrino signal of the
core-collapse supernovae of a 40~M$_\odot$ progenitor. We confirmed the results
of other studies that ${\mu/\tau}$-neutrinos carry an important signature of
the EOS, namely, the temperatures reached inside the proto-neutron star, which
are connected to the stiffness around saturation density. However, the most
prominent EOS aspect related to the neutrino signal of such massive progenitors
is the time until black hole formation. We found that it is dictated by
the global behavior of the EOS and not by single EOS parameters. We showed that
the correlation between time until black hole formation and maximum mass of cold
neutron stars, found in previous EOS studies, does not hold for the extended set
of EOS investigated here. Using the two non-relativistic LS EOS, the collapse to
the black hole takes place systematically earlier than with the RMF EOS, even if
the LS EOS have larger maximum masses for cold neutron stars. This interesting result
was explained by the different response to finite temperature. For the RMF EOS a significantly larger increase of the maximum masses at finite entropy was observed than for the LS EOS.
The states at the beginning of black hole formation in the simulations could be approximated by hydro-static configurations of proto-neutron stars at a constant entropy per baryon of $s=4$~k$_B$. The resulting maximum masses agreed with the results from the simulations. We found a new monotonic correlation between the time until black hole
formation and the maximum masses of these proto-neutron star configurations.
From this correlation and the maximum mass constraint from
\citet{Demorest:etal:2010}, we predict that the neutrino signal of a
core-collapse supernova of a 40~M$_\odot$ progenitor should last at least for
0.5 s starting at bounce, until the black hole forms. If such an event was
measured in the future, the correlation would provide the maximum mass of proto-neutron stars at constant entropy. This would represent a significant constraint for the nuclear EOS at finite entropy, complementary to observations of cold neutron stars. In combination with the existing knowledge about the maximum mass of cold neutron stars this would give information about the temperature dependency of the EOS.

The EOS can also affect the explosion dynamics. In general, soft EOS lead to a
stronger initial shock acceleration where higher matter velocities are obtained
in comparison to stiff EOS \citep{baron85npa,baron85prl,bruenn98}. Therefore
soft EOS result in a more optimistic situation with respect to the initial
energetics at shock breakout shortly after bounce. This was confirmed in the
present study, where the highest matter velocities for the bounce shock were
encountered for the extremely soft LS EOS, and the lowest for the stiff TM1 EOS.
Furthermore, although based on the same RMF parameterization TM1, STOS and HS
have different shock energetics at shock breakout shortly after core bounce, due
to the different description of nuclear matter below saturation density, and the
resulting different deleptonization during the collapse. However, the situation
becomes more similar after the bounce shocks have stalled. The initial dynamical
advantage of a soft EOS is lost with time. It nevertheless leads to a more
extended heating region between the neutrinospheres and the standing bounce
shock. The influence of the EOS on explosions has been investigated in
\citet{MarekJanka:2009} based on simulations in axial symmetry. It would be an
interesting follow-up project to perform a similar EOS comparison, as was done
here, in multi-dimensional simulations, to further tackle the explosion mechanism of
core-collapse supernovae.
\section*{Acknowledgements}
We want to thank R.~K{\"a}ppeli, A.~Arcones, A.~Perego, S.~Typel and
G.~R{\"o}pke for helpful discussions. Furthermore we thank S.~Typel and
A.~Arcones for proofreading the manuscript and their useful comments. We are
also grateful to X.~Roca-Maza, L.~Geng, G.~Lalazissis and P.~Ring for providing
the nuclear mass tables used in the HS EOS. M.H.\ is supported by the High Performance and High Productivity Computing Project (HP2C), and the Swiss National Science Foundation
(SNF) under project number no.\ 200020-132816/1. T.F.\ is funded by
HIC for FAIR project~no.~62800075 and the SNF under project~no.~PBBSP2-133378. J.S.-B.~is supported by the German
Research Foundation (DFG) within the framework of the excellence initiative
through the Heidelberg Graduate School of Fundamental Physics. M.L.\ is funded
by the SNF grant no.~PPOOP2-124879/1. The authors are additionally supported by
CompStar, a research networking program of the European Science Foundation
(ESF). M.L.\ and M.H.\ are also grateful for participating in the EuroGENESIS
collaborative research program of the ESF and the ENSAR/THEXO project.\\

\end{document}